\documentclass[runningheads]{llncs}

\usepackage[T1]{fontenc}
\usepackage{graphicx}
\usepackage{amsmath,amsfonts}
\usepackage{amssymb}
\usepackage{algorithmic}
\usepackage{algorithm}
\usepackage{array}
\usepackage{textcomp}
\usepackage{stfloats}
\usepackage{url}
\usepackage{verbatim}
\usepackage{graphicx}
\usepackage{cite}
\usepackage{booktabs}
\usepackage{hyperref}
\usepackage{tikz}
\usetikzlibrary{shapes.geometric, arrows}
\usepackage{tabularx}
\usepackage[font={small}]{caption}
\usepackage[font={small}]{subcaption}
\usepackage{paralist}
\usepackage{longtable}
\usepackage{listings}
\usepackage{float}

\graphicspath{{figures/}}

\newfloat{lstfloat}{t}{lop}
\floatname{lstfloat}{Listing}

\newcommand{\mypara}[1]{\noindent{\bf {#1}}}

\begin{document}
\title{On the Feasibility of Fingerprinting\\ Collaborative Robot Network Traffic}
\titlerunning{On the Feasibility of Fingerprinting Collaborative Robot Network Traffic}
%

\author{Cheng Tang \and
Diogo Barradas  \and Urs Hengartner \and
 Yue Hu}

\authorrunning{}
%
\institute{University of Waterloo, Waterloo, Canada\\
\email{\{c225tang,dbarrada,urs.hengartner,yue.hu\}@uwaterloo.ca}}

\maketitle              
\begin{abstract}

This study examines privacy risks in collaborative robotics, focusing on the potential for traffic analysis in encrypted robot communications. While previous research has explored low-level command recovery in teleoperation setups, our work investigates high-level motion recovery from script-based control interfaces. We evaluate the efficacy of prominent website fingerprinting techniques (e.g., Tik-Tok, RF) and their limitations in accurately identifying robotic actions due to their inability to capture detailed temporal relationships. To address this, we introduce a traffic classification approach using signal processing techniques, demonstrating high accuracy in action identification and highlighting the vulnerability of encrypted communications to privacy breaches. Additionally, we explore defenses such as packet padding and timing manipulation, revealing the challenges in balancing traffic analysis resistance with network efficiency. Our findings emphasize the need for continued development of practical defenses in robotic privacy and security.

\end{abstract}

\section{Introduction}

In recent years, robotics has rapidly evolved to become an integral part of many industries and domains, including manufacturing, healthcare, transportation, and even personal use, in a fully autonomous fashion or shared-control fashion such as teleoperation~\cite{kebria2018control}.
Among the recent trends in robotic applications, the healthcare and service sectors have seen increased options, thanks to the advent of collaborative robots, i.e., robots intended for close contact with humans~\cite{mihelj2019collaborative}. 
In this context, examples of robotic applications such as surgical assistants~\cite{fiorini2022concepts} bring benefits such as increased dexterity and precision~\cite{francca2020evolution}, and applications in home assistance~\cite{wilson2019robot} guarantee independent living and healthy aging.

However, the integration of robotics in direct interaction with humans also introduces significant privacy concerns~\cite{shah2022can}. The continuous communication between robots and control interfaces, often encrypted yet susceptible to traffic analysis (i.e., the examination of communication metadata like the volume of communication or the frequency of packet exchanges), can enable malicious actors to infer critical information about end users (e.g. a patient's health status, medical procedures being performed, or daily routines in care settings). Such breaches not only pose risks to individual privacy but can also compromise trust in adopting robotic systems. The privacy threat is not just a matter of unauthorized data access; it extends to the potential misuse of sensitive information, which can have far-reaching implications for end-users and service providers~\cite{denning2009spotlight}.

Previous work~\cite{shah2022can} has begun exploring the potential leakage of robotic operations through the analysis of encrypted traffic, focusing on the recovery of low-level commands sent to a robot via the analysis of features such as packet lengths and inter-arrival times. While prior research is tailored to robots controlled by teleoperation interfaces and to the classification of low-level motions via deep learning, we employ signal processing to extract command messages in script-based robot control interfaces.  
By classifying actions based on temporal dependencies via classical machine learning, our model requires few data samples and could potentially be more easily extended across various robotic systems.

We study potential threats arising from traffic analysis in robotic operations, focusing our analysis on the traffic generated by collaborative robots.
We start by leveraging established traffic analysis techniques used in the related domain of website fingerprinting, such as k-FP~\cite{kfingerprinting} or Tik-Tok~\cite{rahman2019tik}, to extract features from our network traces and classify robot actions. However, we find that these methods are unable to account for detailed temporal relationships between sequences of command messages, critical for identifying robotic actions (Section~\ref{sec:known_attacks}). 

To address these shortcomings, we introduce a novel traffic classification approach based on signal processing techniques (Section~\ref{sec:signal_processing_attack}). Our core insight is that distinct robot operation commands generate specific traffic sub-patterns, which can be accurately identified through the application of signal correlation and convolution techniques. 
Further, we develop and evaluate custom traffic analysis defenses for robot operations. Inspired by existing techniques, we implement two metadata obfuscation mechanisms: padding and packet timing manipulation. We analyze the performance overhead of different defense configurations and assess their effectiveness in mitigating traffic analysis. 

In our evaluation, we conduct experiments over a network traffic dataset comprising four robotic actions that were manually generated. The results of our experiments show that our signal processing-based classification technique is able to identify these actions with an accuracy of 97\%, thus revealing that adversaries might be able to infer sensitive robotic actions from encrypted network traffic alone. In turn, our experiments with traffic analysis defenses suggest that balancing traffic analysis resistance with network efficiency is challenging in practical settings, casting the need for additional work towards the development of practical defenses that can be widely applied to the robot operation scenario.

\section{Background}

\subsection{Control Interfaces in Robotics}
\label{subsec:control_methods}

Robotic control interfaces primarily fall into two categories: script-based control and teleoperation, each generating distinct traffic patterns with privacy implications, particularly in encrypted communication settings. Our focus is on script-based control, which is particularly relevant to collaborative robotics.

\mypara{Script-based control.} In this model, robots execute predefined scripts that automate tasks with minimal human intervention. The controller sends structured command messages, and the robot executes them accordingly. For example, a domestic robot may be programmed to open curtains at a set time or sort recyclables from trash. 
These scripts, written in programming languages or proprietary software, define task logic that has evolved from simple sequences to complex behaviors influenced by perception, probabilistic decision-making, and machine learning. Commands are transmitted from the controller to the robot’s onboard computer, which executes actions and generates feedback. This feedback—such as vision, location, or object classification—is sent back to the controller and influences subsequent commands. For instance, in a pick-and-place task, the feedback helps determine grasp adjustments or movement corrections.

\mypara{Teleoperation.} This interface allows human control of robots through real-time mimicry or high-level command execution, but relies on human input for online adjustments. It uses devices like joysticks or VR interfaces to translate human actions into robotic movements, with feedback (e.g., sensory data, video) aiding the operator's situational awareness. Unlike script-based control, where feedback informs automated planning, teleoperation relies on human interpretation.

\mypara{Privacy implications.} Both control modalities produce structured network traffic that can be analyzed to infer robotic tasks or operator intent. In script-based control, adversaries may identify routine procedures by analyzing patterns in command sequences, while in teleoperation, they may infer operator actions based on message timing and frequency. In sensitive applications like healthcare, recurring message patterns could indicate medical routines, while specific sensory data exchanges might reveal patient information or caregiving activities. 
In our work, we focus on fingerprinting the network traffic generated by a specific collaborative robot (the Kinova Gen3 arm, described below), which can autonomously execute a series of high-level commands via script-based control.

\subsection{The Kinova Robotic Arm \& APIs}

Figure~\ref{fig:action-photo} illustrates the Kinova Gen 3 robotic arm, a commercially available robotic arm, accessible to industry, research institutions, and also individual consumers. It is utilized in healthcare and research contexts~\cite{gillini2022dual,colucci2021paquitop,pulikottil2018voice}, providing a rich API with protocols, functions, and libraries for high-level and low-level control. Communication with the Kinova Gen 3 robotic arm is secured using encrypted channels (e.g., TLS) to ensure data integrity and confidentiality.

\mypara{High-level API controls.}
The high-level API abstracts robotic movements, allowing developers to issue complex commands with ease. For example, in a ``pick and place'' task, the robot's controller sends a structured sequence of 
movement commands to navigate to an object, followed by gripper commands to grasp, transport, and release it. These high-level commands generate uniform and consistent network traffic patterns (see Section~\ref{sec:characterization}), as each command maps to a predictable sequence of movements and corresponding data packets.

\mypara{Low-level API controls.}
Low-level controls provide granular access to individual joints, real-time sensor feedback, and custom control algorithms. In tasks like picking and placing, this allows precise wrist and finger adjustments for accurate object handling. Since these controls rely on frequent adjustments and feedback loops, they introduce greater variability in network traffic, as each sensor reading and movement adjustment contributes to a more dynamic packet flow.

\mypara{Privacy implications.}
The interaction between the controller and the Kinova arm generates network traffic patterns that can inadvertently reveal operational details. High-level API controls share similarities with script-based control methods (see Section~\ref{subsec:control_methods}), as both rely on predefined command sequences that shape network traffic patterns. While high-level APIs abstract robot interactions, the potential for traffic analysis and privacy risks remain comparable to those observed in script-based control.
Throughout our study, we focus on fingerprinting the network traffic generated by the Kinova Gen3 arm when controlling it via scripts that rely on the arm's high-level API controls, as these facilitate the analysis of command sequences for abstract and complex tasks.

\subsection{Encrypted Traffic Fingerprinting}

Encrypted traffic fingerprinting denotes a traffic analysis technique based on the examination of the metadata of encrypted traffic~\cite{enisa2019encrypted, al2016adaptive}. By inspecting traffic characteristics like overall communication volume, packet sizes, or inter-packet timing, an eavesdropper can build a profile of the network behavior of some Internet-based application under a given workload (i.e., a traffic \textit{fingerprint}), and then re-identify the occurrence of the same workload by matching its fingerprint.

While we now address earlier efforts aimed at fingerprinting robot traffic, we refer to Section~\ref{sec:rw} for a broader perspective about other traffic fingerprinting contexts (e.g., website~\cite{sirinam2018deep}, video~\cite{beautyBurst}, and IoT~\cite{chowdhury2020network} fingerprinting).

\mypara{Reconstruction of robot operations from encrypted traffic.} Shah et al.~\cite{shah2022can} discuss reconstructing robot high-level movements from lower-level movements in teleoperation scenarios, where an operator directly controls the robot using a movement controller. This approach contrasts with our focus, which lies on script-based applications and reconstructing traffic patterns by analyzing sequences of command messages. Shah et al.'s methodology, which employs a simple neural network with a single hidden layer to analyze encrypted traffic data for identifying robot movements and reconstructing workflows, is tailored to teleoperated systems with a dynamic operator control. This differs from our context of scripted robotic actions, making their approach not directly applicable to our experiments due to the distinct setups involved.

\section{Methodology}

This section describes our evaluation methodology. First, we describe our threat model and our experimental testbed. Then, we detail the metrics used to evaluate the success of attacks and defenses applied to robot control traffic.

\subsection{Threat Model}

\begin{figure}[t]
	\centering
	\includegraphics[width=0.55\linewidth]{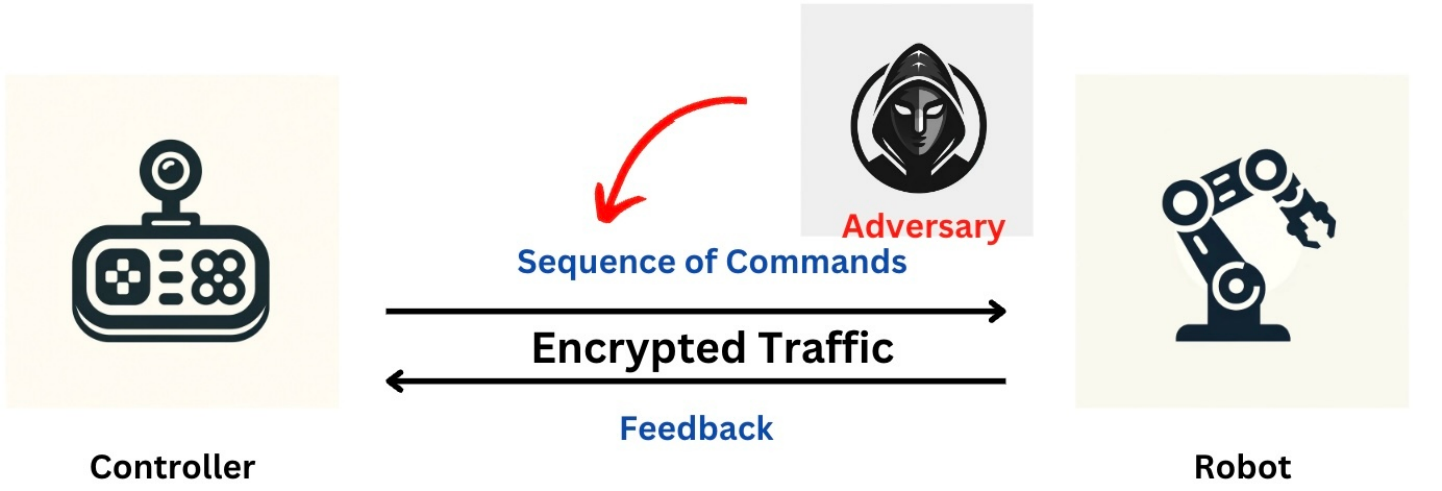}
 \vspace{-0.3cm}
	\caption{Depiction of our threat model.}
        \label{fig:threat-model}
         \vspace{-0.5cm}
\end{figure}

Our threat model is illustrated in Figure~\ref{fig:threat-model}, where an adversary is assumed to passively listen to the communication between a robot and its controller, located in a different network location. As an example, considering a healthcare scenario, the robot could be located in a user's home while being remotely controlled by an operator within a hospital facility. 

\mypara{Fingerprinting setting.} The adversary's goal is to identify the robot’s actions by analyzing traffic patterns generated during control operations, including the exchange of commands and feedback. Beyond recognizing actions, the adversary may also seek to infer fine-grained details, such as the exact commands enabling a given action. These insights could allow near real-time activity prediction.

In general, traffic fingerprinting attacks assume either a \textit{closed}- or \textit{open-world} setting~\cite{sirinam2018deep}, depending on the adversary’s knowledge. In the \textit{closed-world} setting, the adversary knows that the robot is only able to perform a restricted set of actions (monitored set), and aims to identify which of these actions has been performed at a certain time. In the \textit{open-world} setting, the robot is assumed to be able to perform additional actions outside the adversary's monitored set.

Our work assumes the \textit{closed-world} setting, where, despite the set of monitored actions being limited, it is still expected to be sufficient to allow an adversary to infer a robot's activity in specific contexts. For instance, in sensitive healthcare environments, repeated command patterns linked to a concrete set of medical procedures could inadvertently reveal patient-specific information. A sequence of actions unique to a particular therapy may expose a patient’s treatment regimen, while other recurring actions could disclose medication schedules.

\mypara{Adversary's capabilities.} We assume that the adversary has the following main capabilities: First, the adversary has their own robot and controller, whose model/version is identical to the one used by the target user. This allows the adversary to use its own robot (and unrestricted API access) to study traffic patterns. This level of access provides them with a comprehensive understanding of potential traffic patterns, thus offering them a strategic advantage in predicting and interpreting traffic flows. Second, the adversary has the capability to eavesdrop the traffic exchanged between the robot and its controller, e.g., by wiretapping the Internet connections of a given household~\cite{engelberg2022classification}. Third, the adversary can use multiple techniques to analyze traffic flows towards generating features that can help characterize traffic traces (e.g., deriving summary statistics or well-defined patterns through signal processing operations). Finally, the adversary can use machine learning techniques for helping it interpret the command sequences issued to the robot and identify actions issued via the controller.

We also assume that the adversary is limited in its operation. Specifically, the adversary is unable to break the cryptographic primitives used for securing the control channel, and the adversary has no control over the endpoints engaged in communication. Thus, we assume that both the controller and the robotic arm are trusted and remain free from compromises that could help the adversary in collecting additional information about the commands executed by the robot.

\subsection{Experimental Setup}

\mypara{Hardware and software configuration.} Our experiments leverage a Kinova Robotic Arm Gen3~\cite{kinova3}, which remotely receives commands and returns feedback to a workstation acting as a controller. This workstation runs Ubuntu 20.04 LTS and is configured with a 2.40 GHz Intel Core i7-8700T CPU and 32GB RAM.
The workstation and the robotic arm are interconnected using Ethernet cabling to ensure a reliable communication. We use the Kinova API v2.3.0 for controlling the robotic arm and generating command patterns, configuring it to enable TLS-based encrypted communication between the robot and the controller.
To capture the encrypted traffic exchanges between the robot and the controller for our analysis, we use \texttt{\small tcpdump} on the controller workstation.

\begin{figure}[t]
    \centering
    \subfloat[][Pick and place.]{\includegraphics[trim={0cm 0cm 0cm 0cm},clip,width=0.23\linewidth]{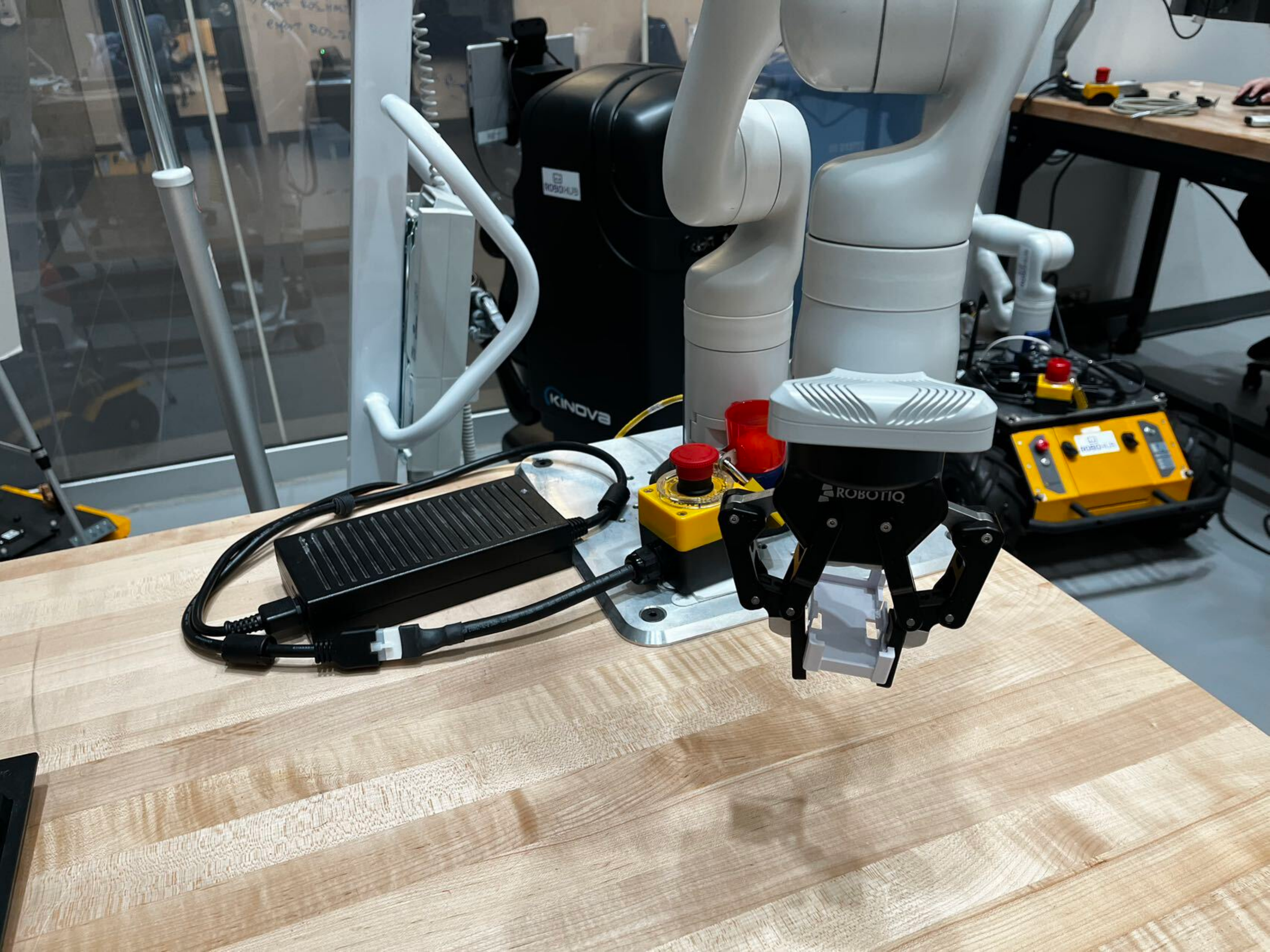}} \hfill
    \subfloat[][Pour water.]{\includegraphics[trim={0cm 0cm 0cm 0cm},clip,width=0.23\linewidth]{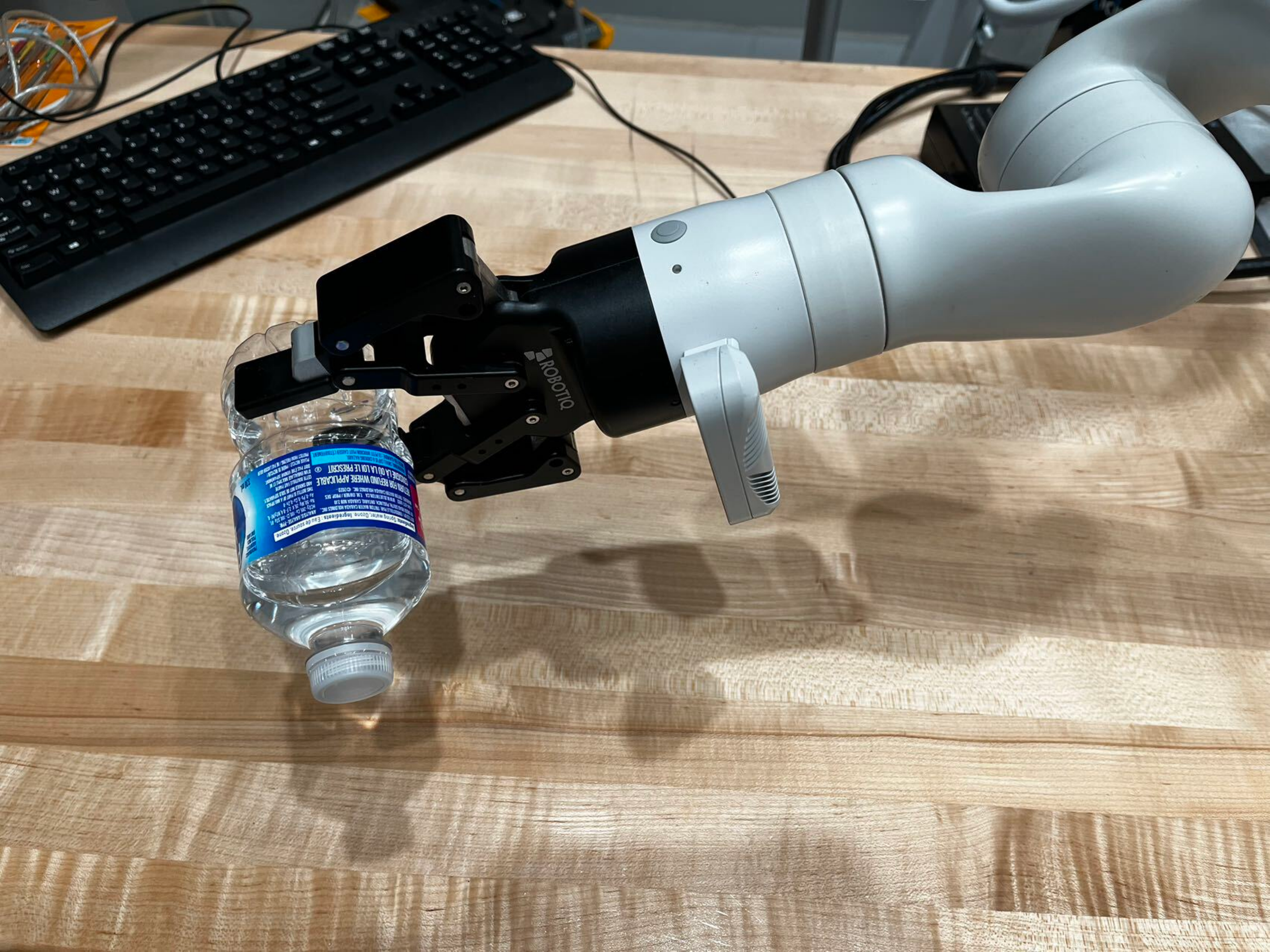}} \hfill
    \subfloat[][Turn on switch.]{\includegraphics[trim={0cm 0cm 0cm 0cm},clip,width=0.23\linewidth]{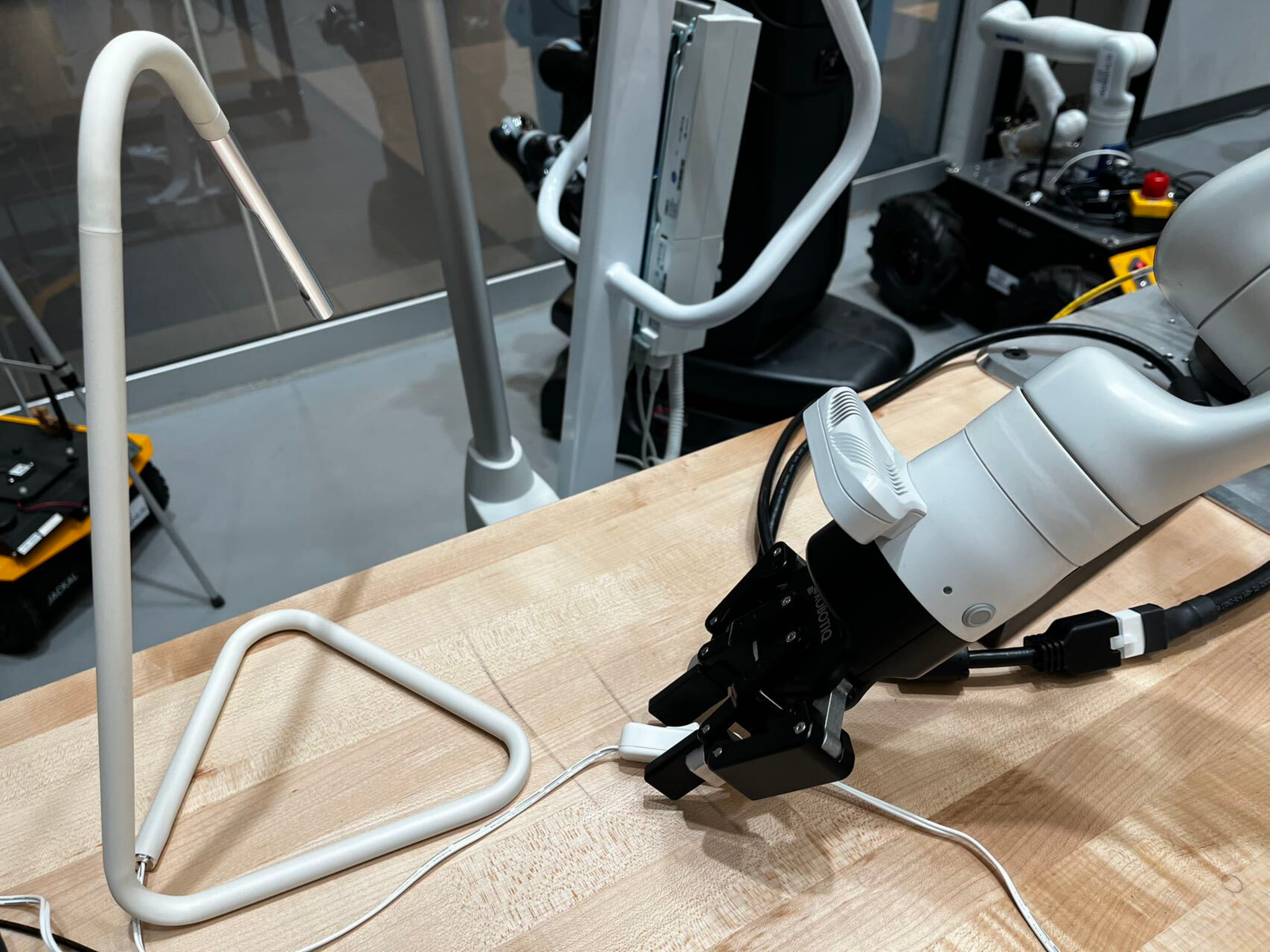}} \hfill
    \subfloat[][Press key.]{\includegraphics[trim={0cm 0cm 0cm 0cm},clip,width=0.23\linewidth]{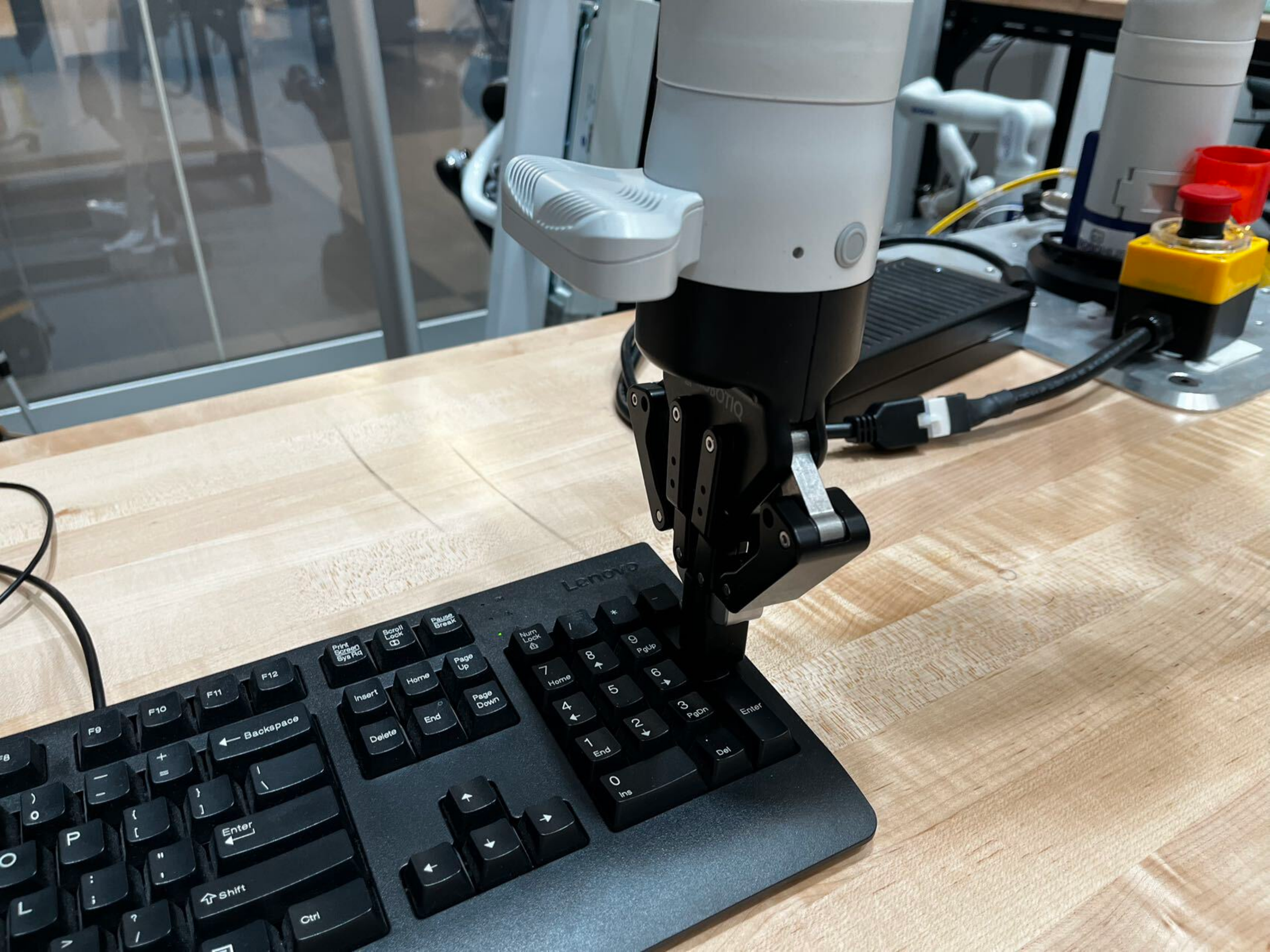}} \hfill
    \vspace{-0.25cm}
    \caption{The Kinova robotic arm executing four actions.}
    \label{fig:action-photo}
    \vspace{-0.6cm}
\end{figure}

\mypara{Robot actions dataset.} We obtain our dataset by collecting the network traces exchanged between the robot and its controller when instructing the robot to execute four distinct actions: a) picking up and placing an object, b) pouring water, c) turning on a switch, and d) pressing a key. Figure~\ref{fig:action-photo} provides a depiction of the robotic arm while performing each of these actions. The actions share a certain degree of similarity; for instance, both pouring water and turning on a switch involve fluid motions and subtle changes in the robot's articulation, which could be reflected in the traffic patterns as similar command message sequences, making it challenging to distinguish between them accurately. The duration of each trace also varies significantly across actions to capture cases where users might take different times to complete the same action. However, all our samples are bounded to a minimum and maximum completion time of 5 and 30 seconds.

We collect a total of 200 samples: 50 for each action. The robot executes actions according to a set of predefined rules that an operator (ourselves) writes in scripts. However, data collection depends on the robot's physical operation, which cannot be easily parallelized without additional robots (unavailable to us). 
The catalog of commands we execute includes different combinations of Cartesian motions (i.e., instructions to move the robot arm in a straight line along the X, Y, and Z axes), the opening/closing of a gripper claw, and adjustments of the gripper's speed. This flexibility allows us to capture a multitude of real-world \textit{trajectories}, where different operators may script the completion of an identical task with a different number of high-level commands or change their order. Each action sample follows a unique trajectory, composed of sub-tasks (e.g., movement, object manipulation) executed at varying points in time. Sub-tasks differ in speed, waypoints, object locations, and task-specific parameters (e.g., pouring angles), enhancing trajectory diversity. We leave the exploration of other approaches to generate command traffic sequences automatically (e.g., generative adversarial networks~\cite{netshare}) to future work.

\mypara{Considered traffic analysis attacks.} For showcasing the potential threats of traffic analysis attacks on robot traffic, we leverage the open-source implementations of attacks used in the context of website fingerprinting (Section~\ref{sec:known_attacks}), including CUMUL~\cite{CUMUL}, k-FP~\cite{kfingerprinting}, Tik-Tok~\cite{rahman2019tik}, and Robust Fingerprinting (RF)~\cite{shen2023subverting}. 
These attacks rely either on manually-engineered traffic features (CUMUL, k-FP), or latent features extracted via deep learning (Tik-Tok, RF).

\mypara{Considered traffic analysis defenses.} We consider two defenses inspired by constant-rate padding approaches~\cite{tamaraw} (Section~\ref{sec:defenses}). The first involves padding individual packets to enforce uniform packet lengths. The second transmits fixed-size packets at a constant rate, potentially including dummy packets, while ensuring compliance with the scheduling constraints required for robot communication.

\mypara{ML models' evaluation.} The  models employed in our study are trained and tested using stratified 10-fold cross-validation to mitigate the impacts of bias in the dataset. We refrain from using an additional validation set since we use all machine learning models with their default hyperparameter configurations.

\subsection{Metrics}

\mypara{Attack evaluation metrics.} We leverage accuracy as our main metric to assess attack performance. Accuracy is defined as the ratio of the number of correctly classified observations to the total number of observations. We also make use of confusion matrices to visualize attack performance across the multiple action classes by inspecting the distribution of correct and incorrect predictions.

\mypara{Defense evaluation metrics.} 
We primarily focus on two aspects: the reduction of an attack's \textit{accuracy} (effectiveness), and the impact of the defense on the robot's performance (efficiency). We measure the efficiency of the defense through the \textit{bandwidth utilization} and \textit{latency increase} experienced by the robot's activities. To compute bandwidth utilization overheads, we compare the amount of additional data exchanged due to the defense. To compute latency impacts, we compute the delays introduced in the communication between the robot and the controller, and assess the potential impacts of such delays on the correct operation of the robot and its ability to complete a designated task.

\section{Characterization of Robot Actions}
\label{sec:characterization}

In this section, we characterize the actions composing our dataset, showcasing both the inter-class and intra-class variability of the generated traffic traces. 

\begin{figure}[t]
	\centering
    \subfloat[Cartesian movement (sample 1).]{\includegraphics[trim={0.3cm 0.3cm 0.3cm 0.3cm},clip,width=0.26\linewidth]{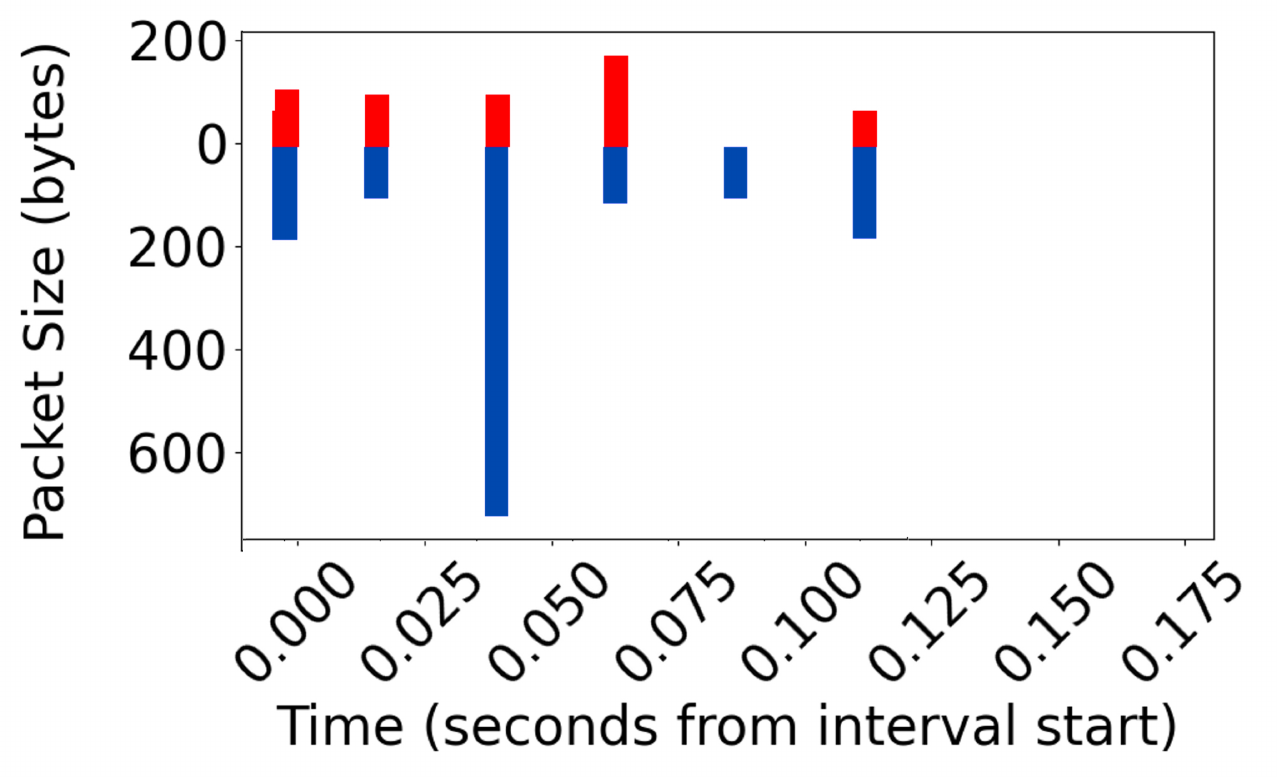}} \hfill
	\subfloat[Cartesian movement (sample 2).]{\includegraphics[trim={2cm 0.3cm 0.3cm 0.3cm},clip,width=0.24\linewidth]{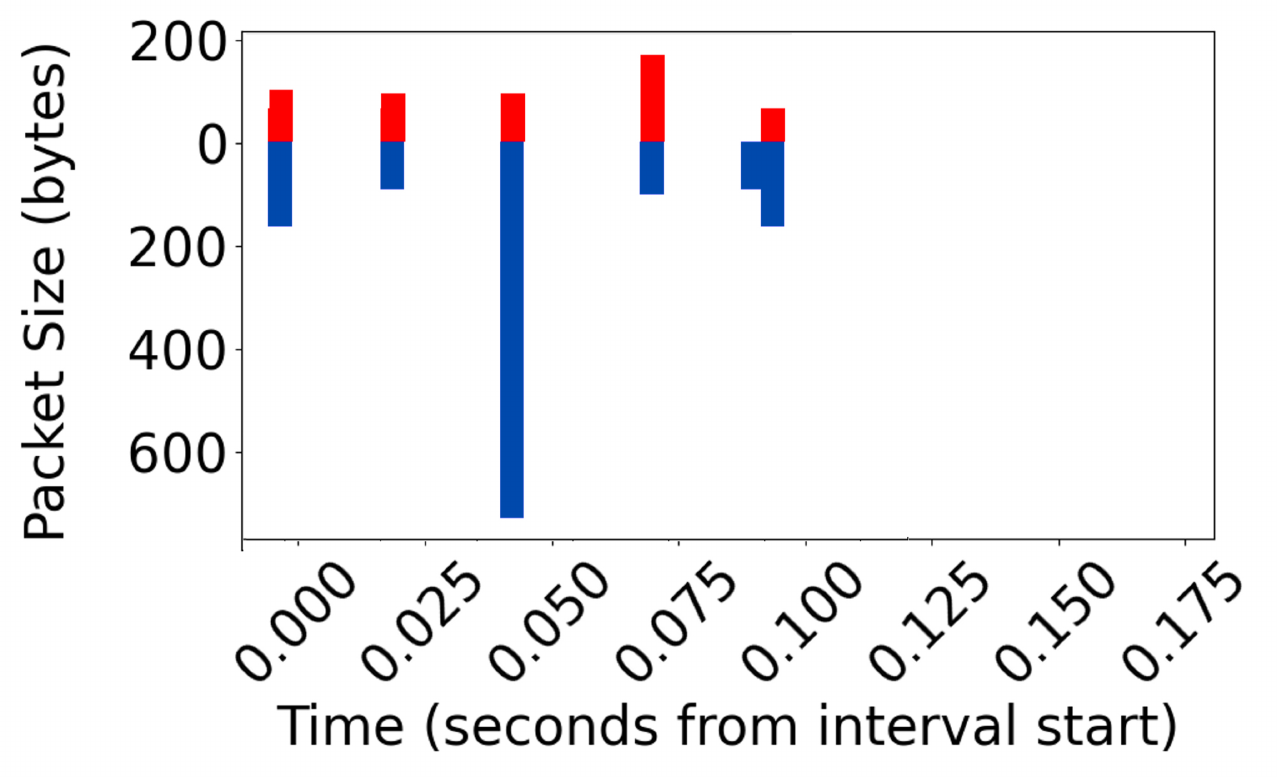}\label{fig:traffic_trace_for_a_Cartesian_Command_for_action_sample2}} \hfill
	\subfloat[Move gripper (sample 1).]{\includegraphics[trim={4cm 0.3cm 0.3cm 0.3cm},clip,width=0.24\linewidth]{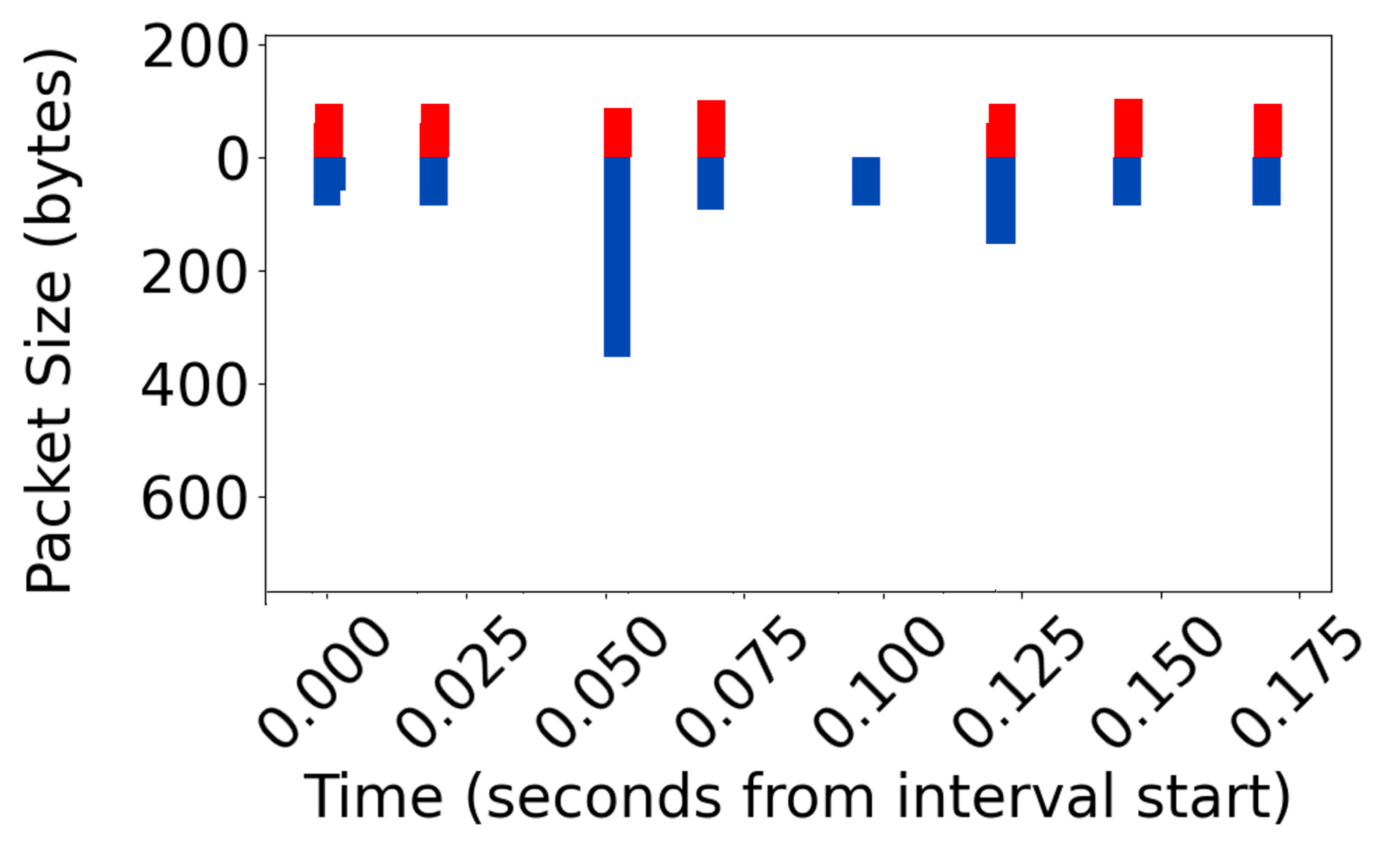}\label{fig:traffic_trace_for_a_Gripper_Position_Command1}} \hfill
	\subfloat[Move gripper (sample 2).]{\includegraphics[trim={4cm 0.3cm 0.3cm 0.3cm},clip,width=0.24\linewidth]{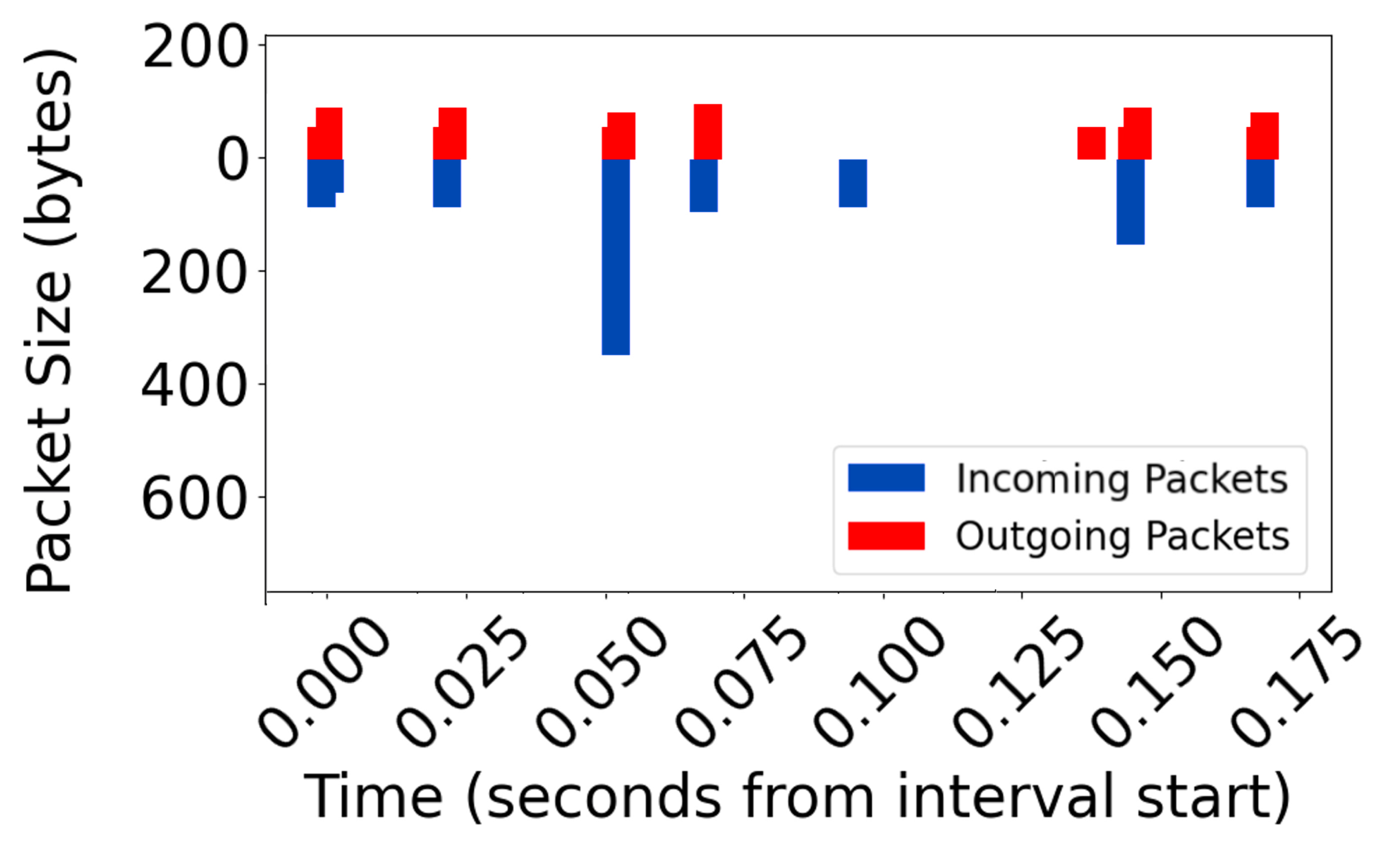}\label{fig:traffic_trace_for_a_Gripper_Position_Command2}}
 \vspace{-0.2cm}
	\caption{Network traces for two kinds of commands.}
	\label{fig:traffic_traces_commands}
 \vspace{-0.7cm}
\end{figure}

\mypara{Individual commands have stable traffic signatures.}  Figure~\ref{fig:traffic_traces_commands} shows two different network traces for a Cartesian command and a gripper movement command, respectively. We can see that different executions of the same command result in similar traffic patterns, but that these patterns also differ amongst each different kind of command. In more detail, Cartesian commands involve moving the robot arm in three-dimensional space and typically span a shorter period of time compared to gripper position commands. These Cartesian movements usually elicit a longer feedback packet from the robot arm, indicative of the arm’s positional adjustments in a 3D space. In contrast, gripper position command messages control the opening and closing of the robot's gripper, and consist of packets just over 100 bytes in either direction. Each of these packets corresponds to an adjustment in the gripper's position.

\begin{figure}[b]
\vspace{-0.5cm}
	\centering
    \subfloat[][Pick and place.]{\includegraphics[trim={0.1cm 0.1cm 0.7cm 0.1cm},clip,width=0.26\linewidth]{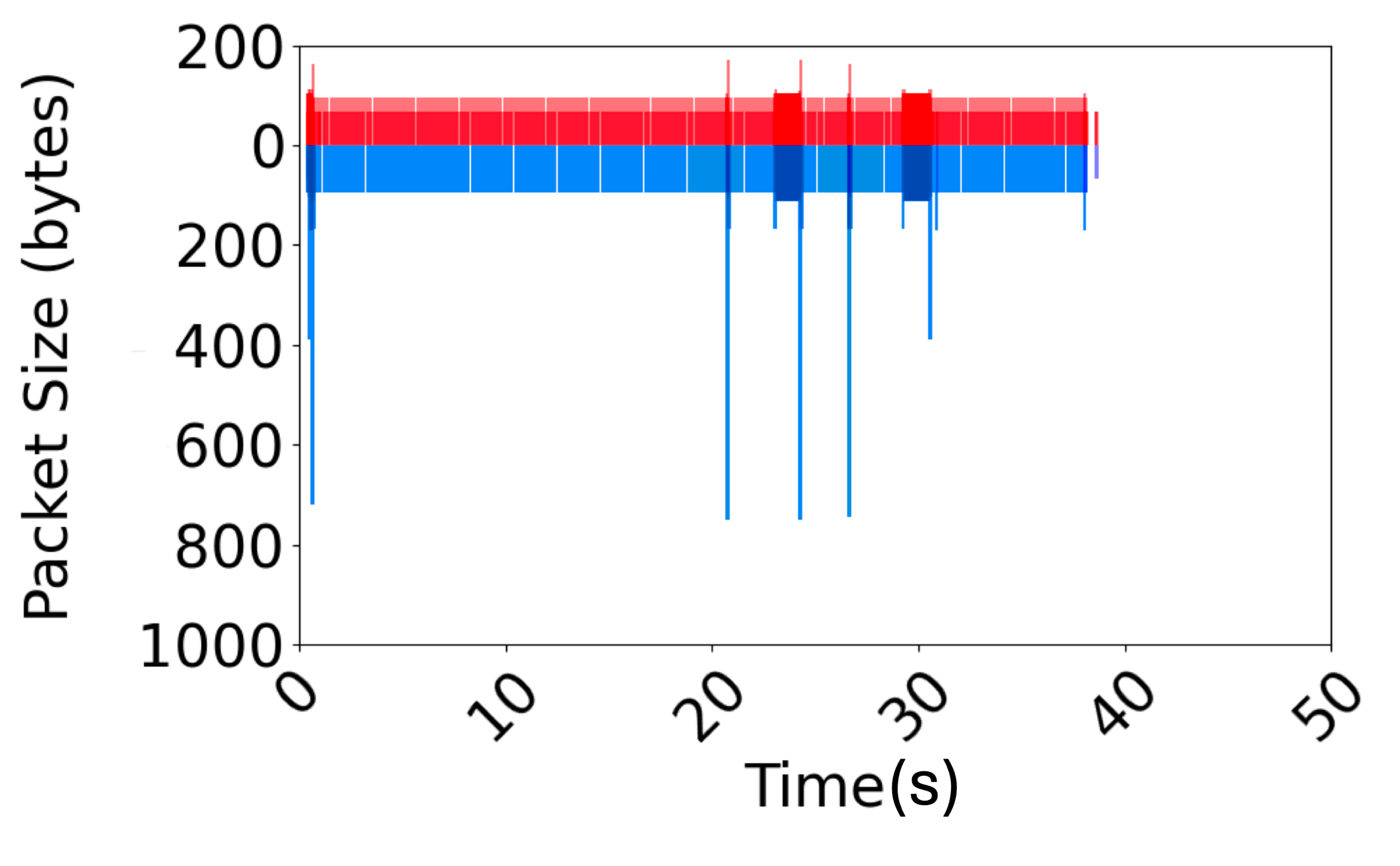}} \hfill
	\subfloat[][Pour water.]{\includegraphics[trim={4.5cm 0.1cm 0.1cm 0.1cm},clip,width=0.24\linewidth]{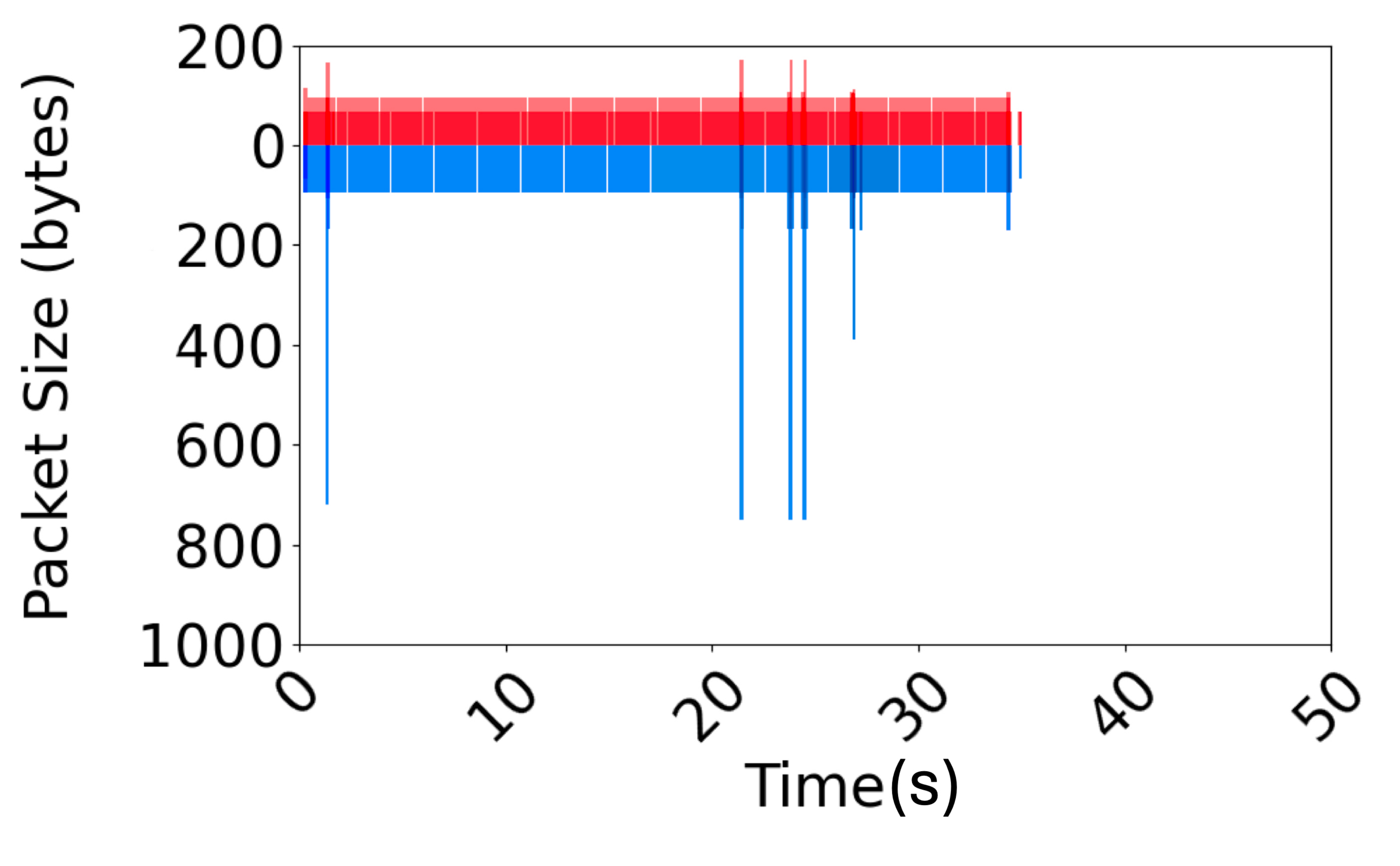}} \hfill
 \subfloat[][Turn on switch.]{\includegraphics[trim={4.5cm 0.1cm 0.1cm 0.1cm},clip,width=0.24\linewidth]{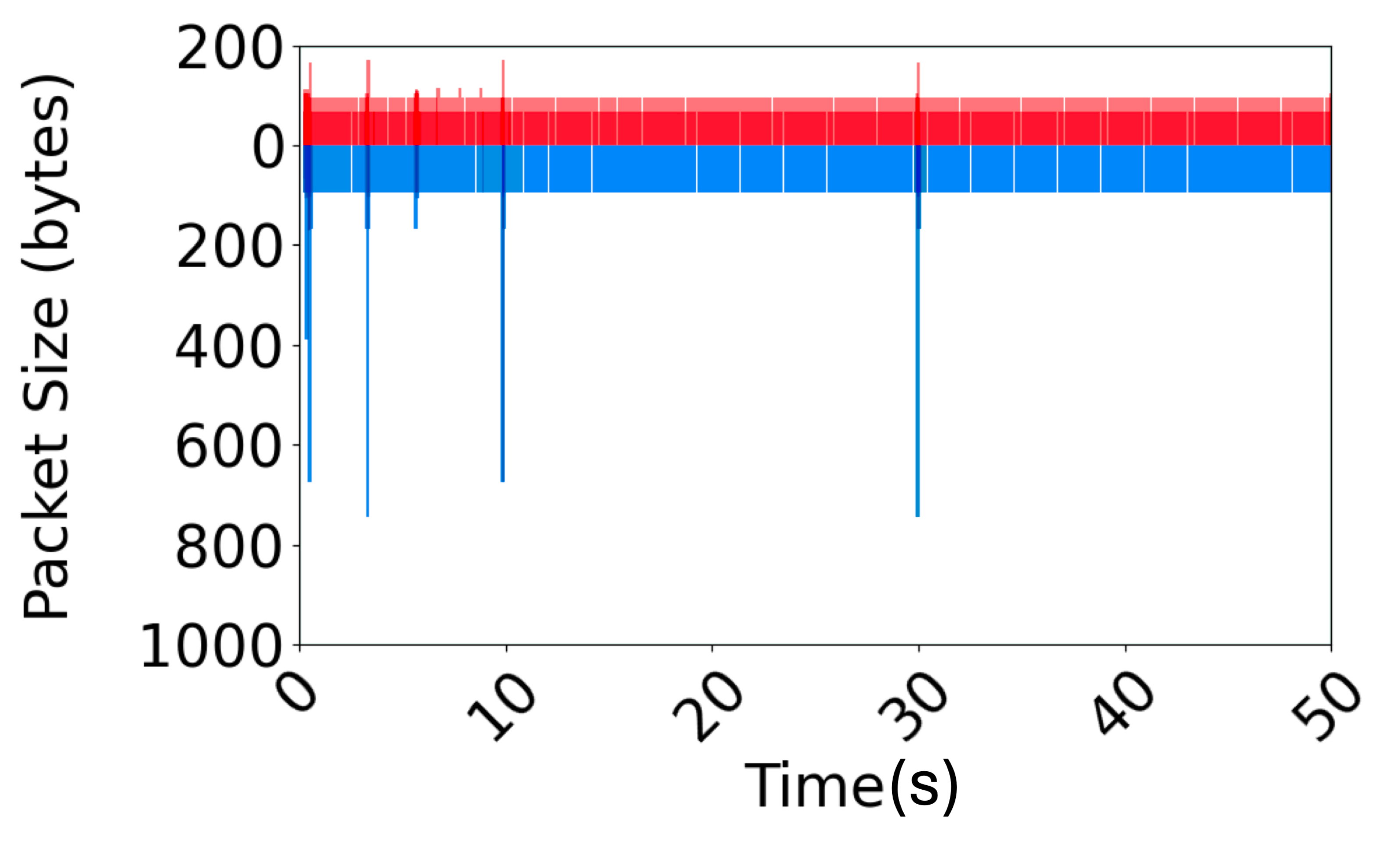}}\hfill
     \subfloat[][Press key.]{\includegraphics[trim={4.5cm 0.1cm 0.1cm 0.1cm},clip,width=0.24\linewidth]{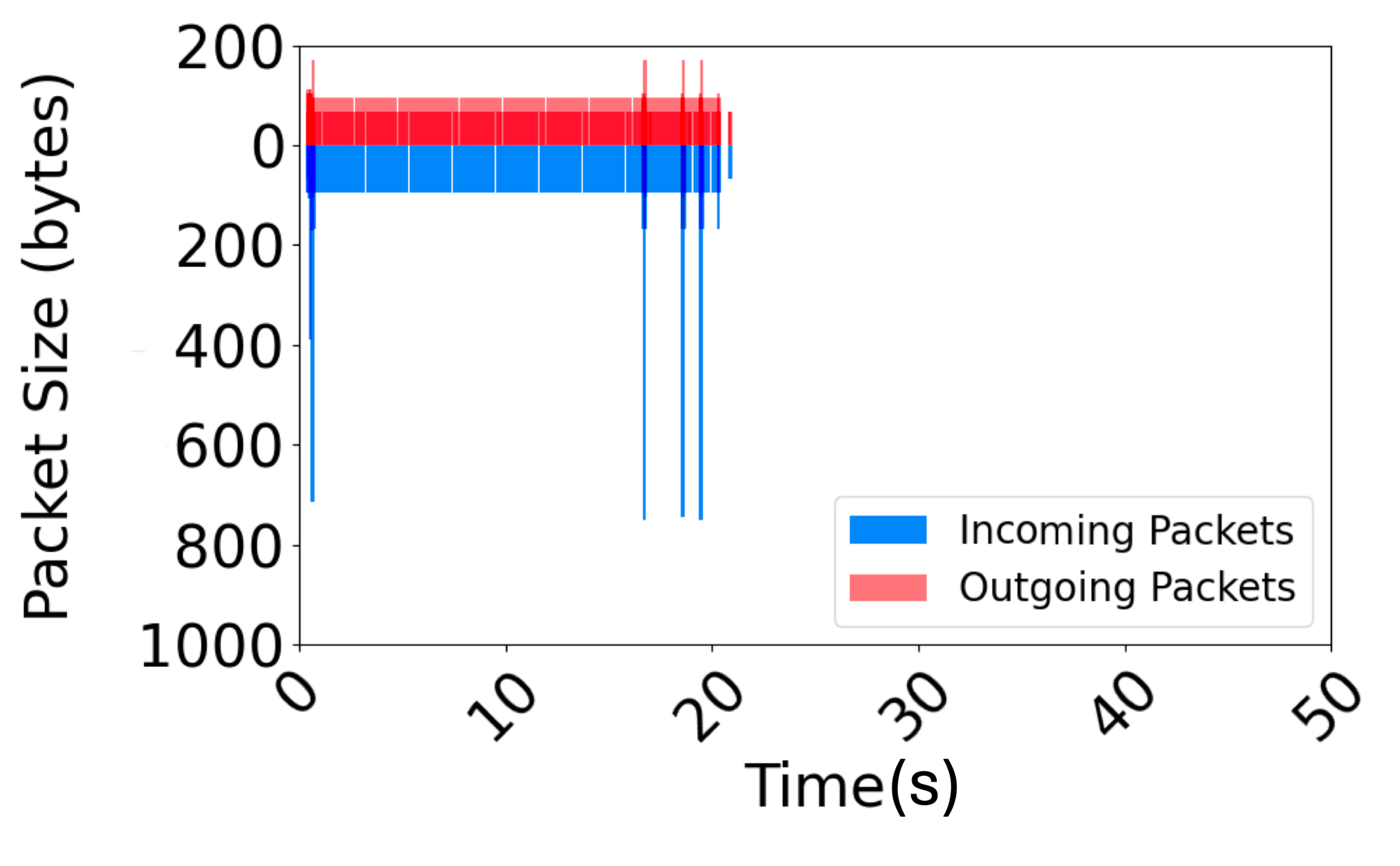}} 
     \vspace{-0.2cm}
	\caption{Examples of robot actions' packet lengths over time.}
	\label{fig:size-time-plot}
\end{figure}

\mypara{Different actions result in different traffic patterns.} Figure~\ref{fig:size-time-plot} shows example traffic patterns for the four classes of actions included in our dataset, showcasing both incoming and outgoing traffic from the point of view of the controller workstation. At a high level, we can see that each different action generates a disparate traffic pattern. For instance, the pick and place action (Figure~\ref{fig:action-photo}a) generally involves at least two distinct gripper command messages, before and after the act of picking and placing an object. This is reflected in the traffic as two dense clusters of gripper position or speed command messages, indicating the gripper's motion to open and close. Despite the two actions being markedly dissimilar, the traces generated by the pour water action (Figure~\ref{fig:action-photo}b) exhibit some similarities tied to specific commands that instruct the closing of the gripper to grab the bottle and opening it after pouring.

In turn, actions such as turning on a switch (Figure~\ref{fig:action-photo}c) or pressing a key (Figure~\ref{fig:action-photo}d) typically involve a tapping motion, which requires the robot to close the gripper beforehand. These actions do not necessitate additional gripper motions afterwards, although the gripper may open if required by subsequent tasks. The timing between Cartesian command messages also offers intricate details; tapping motions are usually swift, leading to shorter intervals between command messages. 
Other noteworthy detail includes the observation of interleaved occurrence of different commands. For instance, in the pick and place action, we can observe the presence of Cartesian command message between two gripper command messages, as seen in Figure~\ref{fig:size-time-plot}c around the 25 second mark. These temporal dependencies that are characteristic to each action may provide actionable information for building an effective classifier.

\begin{figure}[t]
	\centering
	\hfill \subfloat[][Sample 1.]{\includegraphics[trim={0.1cm 0.8cm 0.1cm 0.1cm},clip,width=0.3\linewidth]{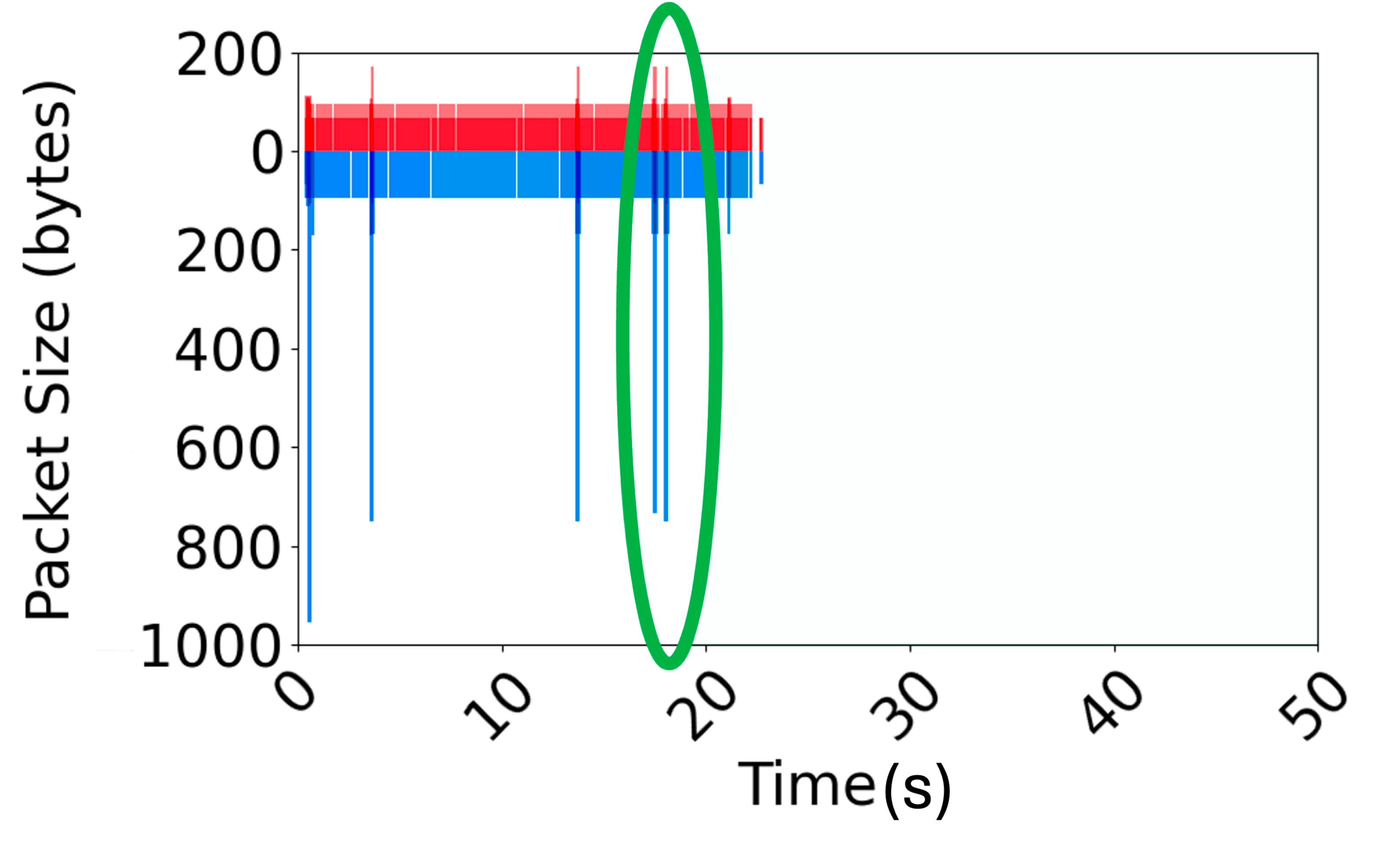}\label{fig:traffic-trace-turn-on-tap-sample1}} \hfill
	\subfloat[][Sample 2.]{\includegraphics[trim={0.1cm 0.8cm 0.1cm 0.1cm},clip,width=0.3\linewidth]{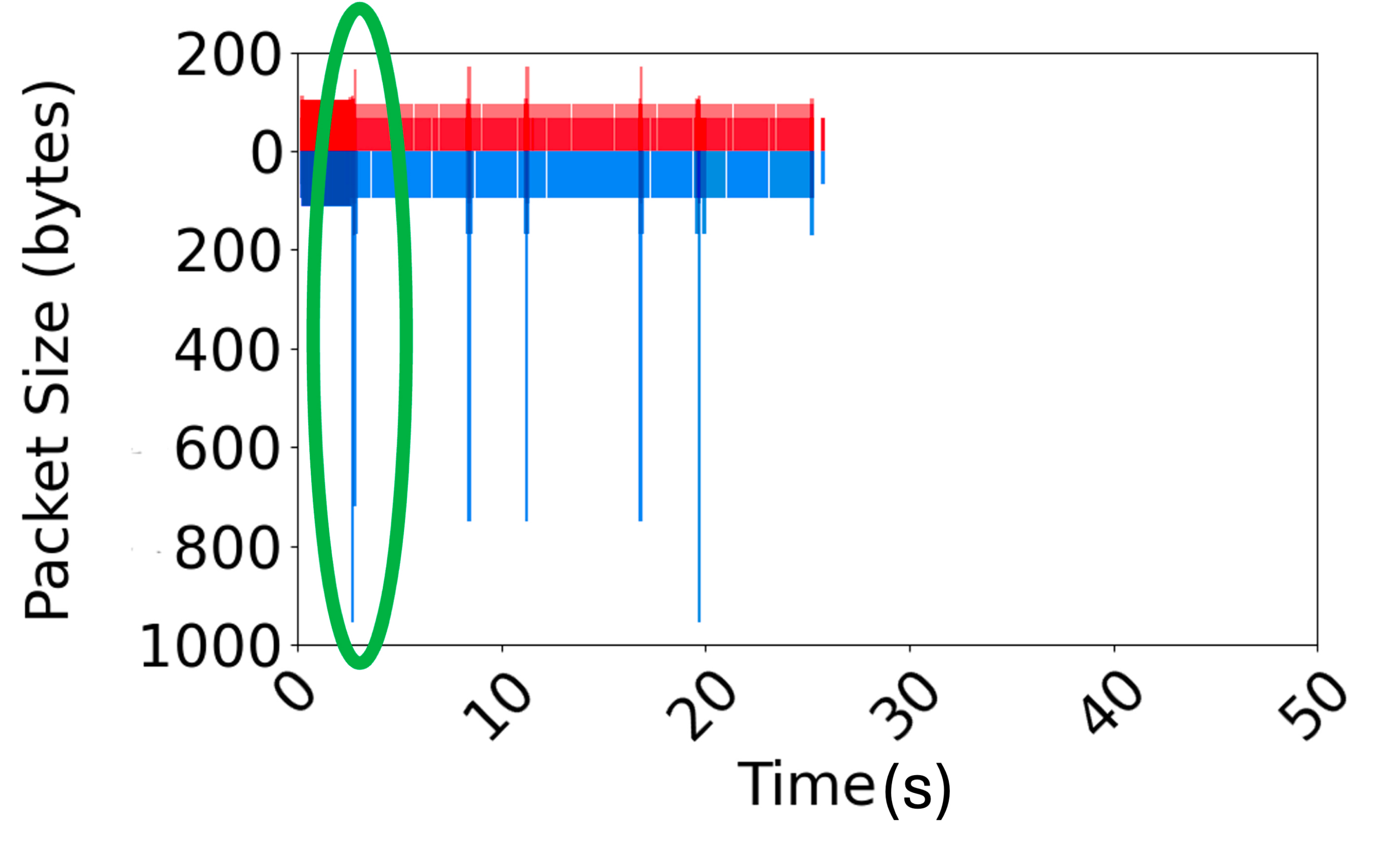}\label{fig:traffic-trace-turn-on-tap-sample2}} \hspace{0.15\linewidth}
 \vspace{-0.3cm}
	\caption{Examples of two ``turn on switch'' network traffic traces.}
	\label{fig:size-time-plot-comparison}
 \vspace{-0.5cm}
\end{figure}

\mypara{Traffic patterns for a given action are highly variable.} Figure~\ref{fig:size-time-plot-comparison} presents two variations of the turn on switch action, showcasing that the traffic patterns for a given action can also be distinct and challenging to recognize by simple observation. However, upon closer inspection, the composition of individual commands sent to the robot can help us identify an action, such as the short period of time between two Cartesian command messages, as circled in green in Figure~\ref{fig:size-time-plot-comparison}a, and a gripper command (closing the gripper) at the beginning of the action to prepare for the tapping motion, immediately followed by a Cartesian command message, as circled in green in Figure~\ref{fig:size-time-plot-comparison}b.

The results of our characterization suggest that there are multiple sources of information in robot traces that may be leveraged for enabling their accurate classification. Next, we experiment with different traffic classification approaches and assess whether these enable the successful identification of robot actions.

\section{Exploiting Known Attacks for Action Identification}
\label{sec:known_attacks}

Multiple classifiers have been proposed for traffic fingerprinting, with website fingerprinting being one of the most well-studied contexts. In this section, we outline how popular classifiers operate, explain their feature selection rationale, and evaluate their performance when fingerprinting robot operation traffic.

\subsection{ML-based Website Fingerprinting Attacks}

We now describe four prominent classifiers that have been used for successful traffic analysis attacks in the context of website fingerprinting. 

\mypara{CUMUL~\cite{CUMUL}.} The feature set used in this attack includes the total number of incoming and outgoing packets, total bandwidth used in each direction. Additionally, 100 features are generated from the cumulative sum of packets' sizes of the connection at different points in a network trace. The attack leverages a Support Vector Machine with an RBF kernel.

\mypara{k-FP~\cite{kfingerprinting}.} This attack introduces a combination of features used in previous attacks with novel traffic characteristics, leading to a systematic analysis of 150 traffic features. The classifier works by building a fingerprint for each web page using a modification of the Random Forest algorithm. Then, the attack employs the k-Nearest Neighbors classifier to predict web page accesses.

\mypara{Tik-Tok~\cite{rahman2019tik}.} This attack builds-up on Deep Fingerprinting (DF)~\cite{sirinam2018deep} and is based on a deep convolutional neural network (CNN) that extracts latent features from network traces to classify websites. Its input is based on a directional-timing representation of traffic, obtained by multiplying a packet's direction (incoming/outgoing represented as -1/+1) with its inter-packet arrival time.

\mypara{Robust Fingerprinting (RF)~\cite{shen2023subverting}.} This attack introduces a Traffic Aggregation Matrix (TAM), which divides network traffic into fixed-size time windows or a sequence of N packets, with each row corresponding to an individual packet and columns representing extracted features of that packet. It identifies website-specific traffic signatures by leveraging TAM's spatial structure to extract hierarchical patterns using a CNN-based classifier.

\subsection{Attacking Robot Operation Traffic}

We now present our main findings after applying the classifiers introduced above for attempting the identification of robot actions issued via the controller.

\begin{figure}[t]
    \centering
    \subfloat[][CUMUL.]{\includegraphics[trim={0.1cm 0.1cm 0.1cm 0.1cm},clip,height=2.7cm]{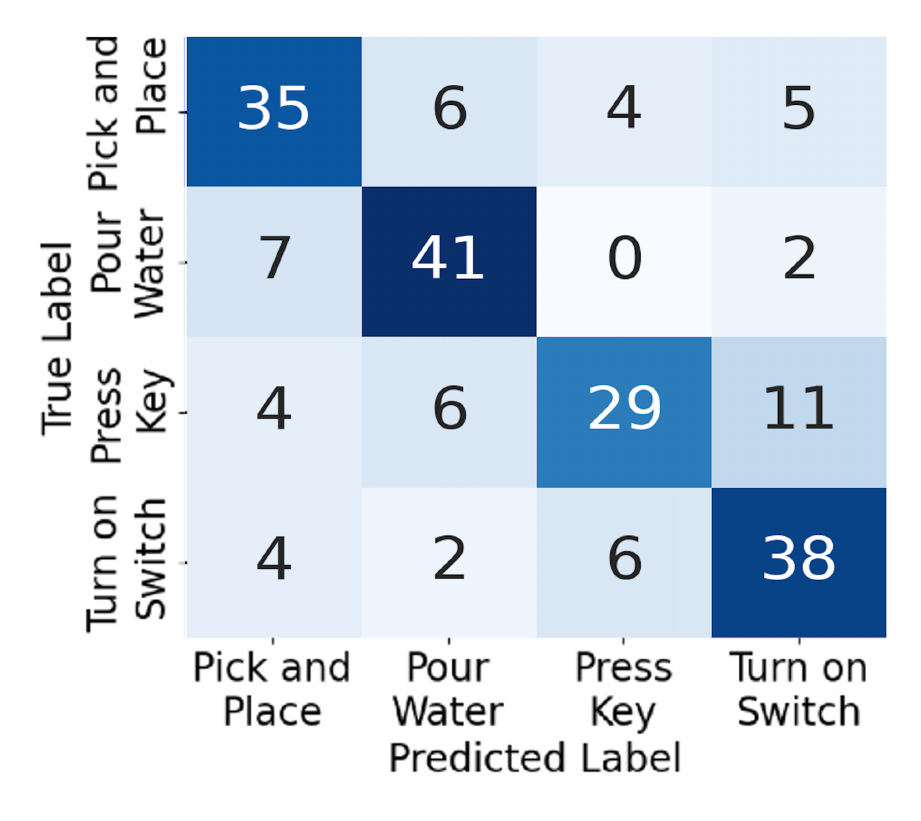}\label{fig:cumul_cm}} \hfill
    \subfloat[][K-FP.]{\includegraphics[trim={0.1cm 0.1cm 0.1cm 0.1cm},clip,height=2.7cm]{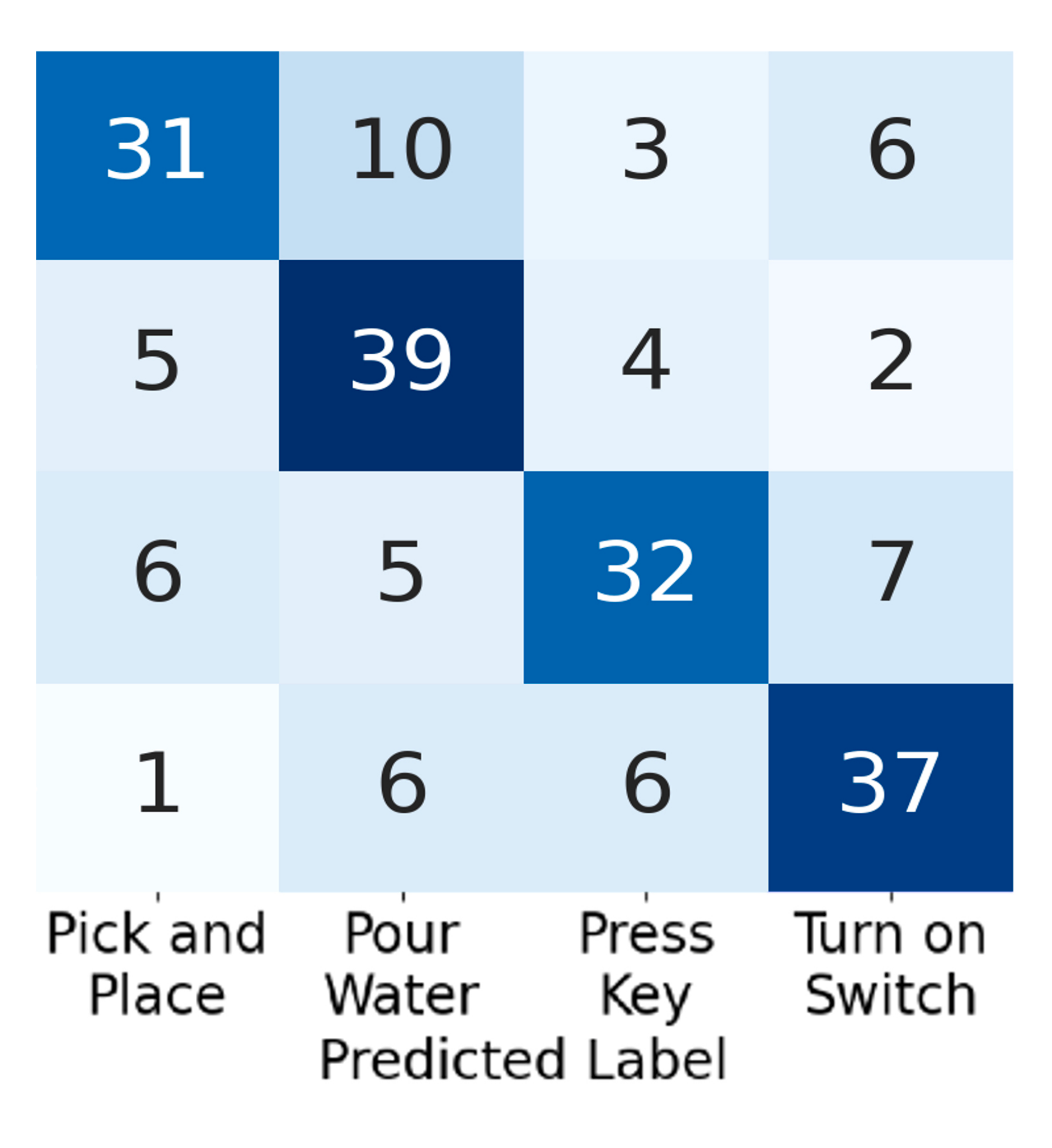}\label{fig:kfp_cm}} \hfill
    \subfloat[][Tik-Tok.]{\includegraphics[trim={0.1cm 0.1cm 0.1cm 0.1cm},clip,height=2.7cm]{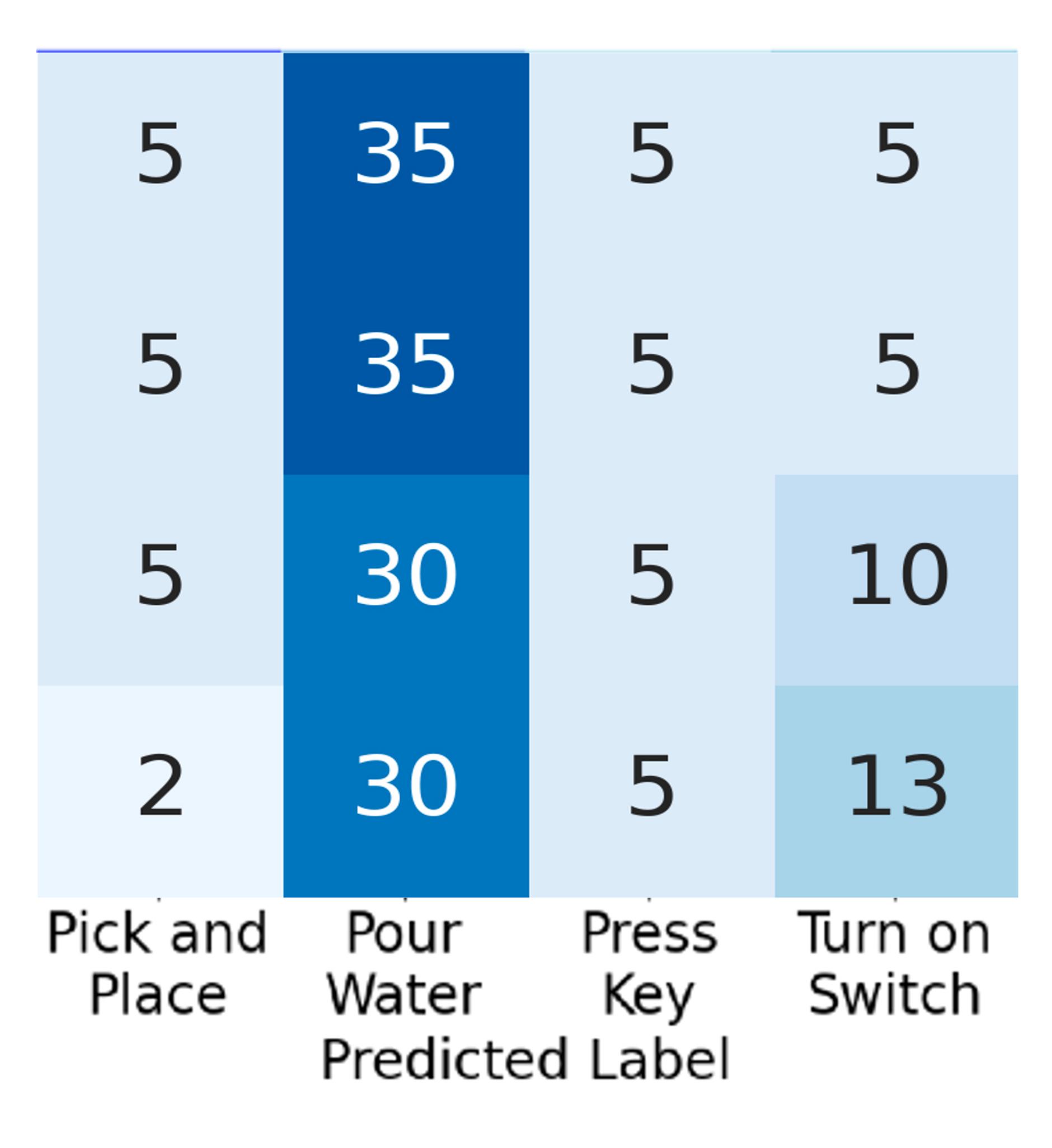}\label{fig:tiktok_cm}} \hfill
    \subfloat[][RF.]{\includegraphics[trim={0.1cm 0.1cm 0.1cm 0.1cm},clip,height=2.7cm]{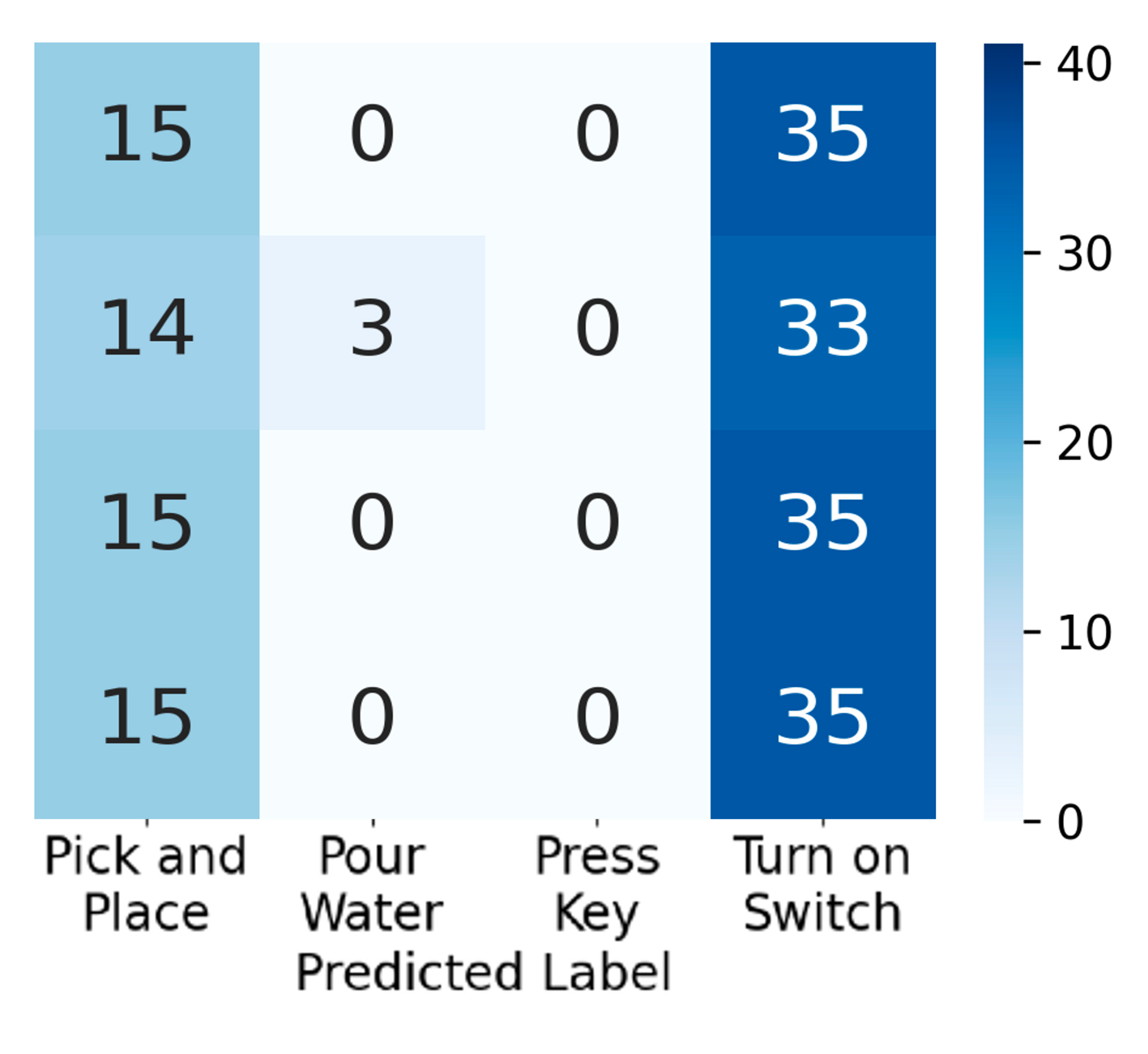}\label{fig:rf_cm}}
     \hfill
    \vspace{-0.2cm}
    \caption{Confusion matrices for website fingerprinting attacks deployed on robot traffic.}
    \label{fig:existing-confusion-matrices}
    \vspace{-0.5cm}
\end{figure}

\mypara{Existing website fingerprinting attacks fail to identify robot actions.} Our experiments reveal that the attacks introduced in the previous section obtain a poor accuracy in identifying the robotic actions in our dataset. Specifically, k-FP and CUMUL achieve accuracies of 69.5\% and 71.5\%, respectively, while models relying on deep learning, such as RF, struggle with limited data, performing only marginally better than random guessing. In particular, the Tik-Tok attack reaches just 29.0\% accuracy, while RF attains 26.5\%. Figure \ref{fig:existing-confusion-matrices} illustrates the confusion matrices for each classifier.

\mypara{Why do these attacks fail?}
The relatively low performance of these attacks in identifying robot actions can be attributed to a set of fundamental differences in the nature of robotics traffic compared to web traffic. As we observed in Section~\ref{sec:characterization}, the robotic actions considered in our work generate highly variable traffic patterns with intricate temporal dependencies  that may not be typically present in web traffic. 
Among these, we highlight nuanced variations in gripper speed and position commands, as well as specific timing intervals between Cartesian command messages. Our results suggest that these dynamic characteristics are not effectively captured by traditional website fingerprinting classifiers, failing to adequately capture local information in a trace, for example, the exact timing at which specific robot operation commands are placed.

\mypara{What can we do about it?} We argue that the presence of command-specific patterns in robotics traffic, such as visible instances of gripper command messages or distinct intervals between Cartesian commands, demands a more specialized approach for accurate classification. These patterns exhibit structured temporal dependencies that standard classifiers designed for web traffic struggle to capture effectively. 
Next, we depart from the use of established classifiers used for fingerprinting network traffic, and introduce a new fingerprinting approach based on signal processing techniques, which allow us to better analyze temporal structures, extracting features inherent to the different commands composing robotic actions without high data requirements and computational cost.

\section{Signal Processing-based Robot Action Identification}
\label{sec:signal_processing_attack}

\begin{figure*}[t!]
	\centering
	\includegraphics[width=0.95\linewidth]{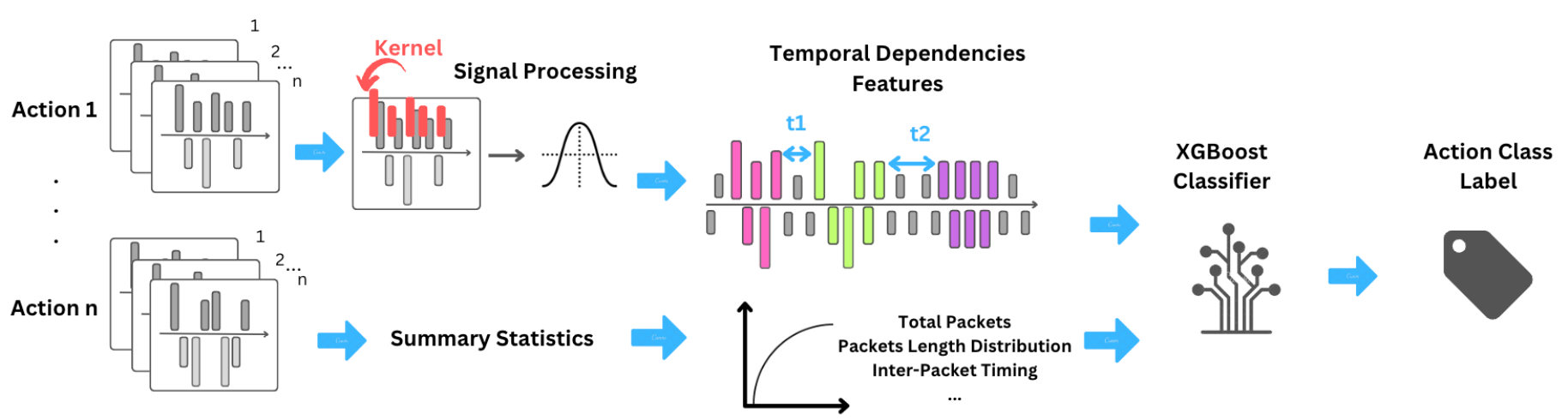}
 \vspace{-0.3cm}
	\caption{\label{fig:analysis-pipeline}Overview of our signal processing-based traffic analysis pipeline.}
 \vspace{-0.3cm}
\end{figure*}

In this section, we introduce a new signal processing-based method to detect network events, and explore whether an adversary could use these methods to enhance robotic action identification capabilities. Specifically, we focus on detecting network events tied to particular robot commands, which serve as the building blocks for constructing action-specific traffic fingerprints.

\mypara{Traffic analysis pipeline overview.} Figure~\ref{fig:analysis-pipeline} depicts a bird's-eye view of our analysis pipeline. Network traces representing different actions start by undergoing a set of signal processing operations (correlation and convolution, see Section~\ref{subsec:pattern-matching}) whose results we rely on to build a set of temporal dependency features. These features are then combined with a set of more generic summary statistics obtained from the network traces (inspired in those used by the ML classifiers analyzed in Section~\ref{sec:known_attacks}). Finally, we use these features to train (and later test the success of) an XGBoost~\cite{chen2016xgboost} classifier geared at identifying the actions performed by the robot. We employ XGBoost given that it is known to outperform other classical machine learning models in both computational speed and model performance, and has been successfully used for traffic analysis~\cite{usenix-mpt}.

Next, we describe our signal processing approach, detail how it is able to recognize temporal dependencies and statistical features of robot control traffic, and show how our method achieves a high accuracy in robot action identification.

\subsection{Pattern Matching Operations}
\label{subsec:pattern-matching}

Our traffic analysis methodology places a large emphasis on the recognition of traffic patterns by leveraging two basic mathematical operations: \textit{convolution} and \textit{correlation}. Next, we provide the necessary background on these operations and then address how their results can be composed into a set of temporal dependencies and statistical features that help us build an accurate classifier.

\mypara{Convolution.} Convolution is a mathematical operation that combines two signals to produce a third, which reflects how the shape of a signal is modified by the other. For two discrete signals \( x[n] \) and \( h[n] \), their convolution is given by:

\vspace{-0.2cm}
{\footnotesize
\[
(x \ast h)[n] = \sum_{m=-\infty}^{\infty} x[m] \cdot h[n-m]
\]
}
\vspace{-0.2cm}

Here, \( h[n] \) can be thought of as a ``kernel'' or ``filter'' that is slid across \( x[n] \). In the realm of traffic pattern recognition, if \( x[n] \) is the observed traffic and \( h[n] \) is a known pattern, then a significant spike in \( (x \ast h)[n] \) at a particular \( n \) indicates the presence of the known pattern in the observed traffic at that point.

\mypara{Correlation coefficient.} The correlation coefficient, often denoted \( r \), quantifies the linear relationship between two signals. For two discrete signals \( x[n] \) and \( y[n] \), their correlation at lag \( m \) is:

\vspace{-0.2cm}
{\footnotesize
\[
r[m] = \sum_{n=-\infty}^{\infty} x[n] \cdot y[n+m]
\]
}
\vspace{-0.2cm}

The value of \( r[m] \) peaks when \( x[n] \) and \( y[n+m] \) align most closely. In the context of traffic analysis, \( x[n] \) could be a known command pattern and \( y[n] \) the observed traffic pattern. A peak in \( r[m] \) indicates the presence of the command pattern within the observed traffic at lag \( m \). In addition to identifying command patterns, correlation coefficients have been used in other network traffic analysis tasks such as anomaly detection~\cite{5341523} or traffic correlation~\cite{nasr2017compressive}.

In our experiments, we find that both correlation and convolution play an important role in detecting repetitive patterns within encrypted traffic. While the operations are closely related (they are in fact identical for symmetric \( h[n] \)), they offer different interpretations and can be used/combined based on the specific requirements of the traffic analysis task at hand.

\subsection{Signal Processing-based Traffic Features} 

In our analysis, we utilize both convolution and correlation-based techniques to extract information on different types of command messages. Some command messages are one-time-only patterns, where convolution is particularly effective. This method allows us to highlight these singular patterns within the broader traffic data. In contrast, other commands are repetitive and may have varying duration; these are better captured using the correlation coefficient to reduce noise and false positives. For instance, we use the correlation coefficient for detecting command messages like gripper speed, where patterns are recurring. Our work primarily focuses on three types of command messages: Cartesian movements, gripper movements, and gripper speed adjustments. Each type presents unique pattern characteristics in network traffic. When introducing a new command message type for detection, we explicitly assess the nature of the command's traffic pattern. This assessment guides our decision on whether to apply convolution — best for one-time, distinct patterns — or the correlation coefficient, which excels in identifying and analyzing recurring patterns within the traffic.

Next, we detail two sets of traffic statistics we extract via convolution and correlation-based methods, and then detail how we combine them to produce the final feature set used by our classifier.

\begin{figure}[t]
    \centering
    \begin{subfigure}[t]{0.48\linewidth}
        \centering
        \includegraphics[trim={0.1cm 1cm 0.1cm 0.1cm},clip, width=0.53\linewidth]{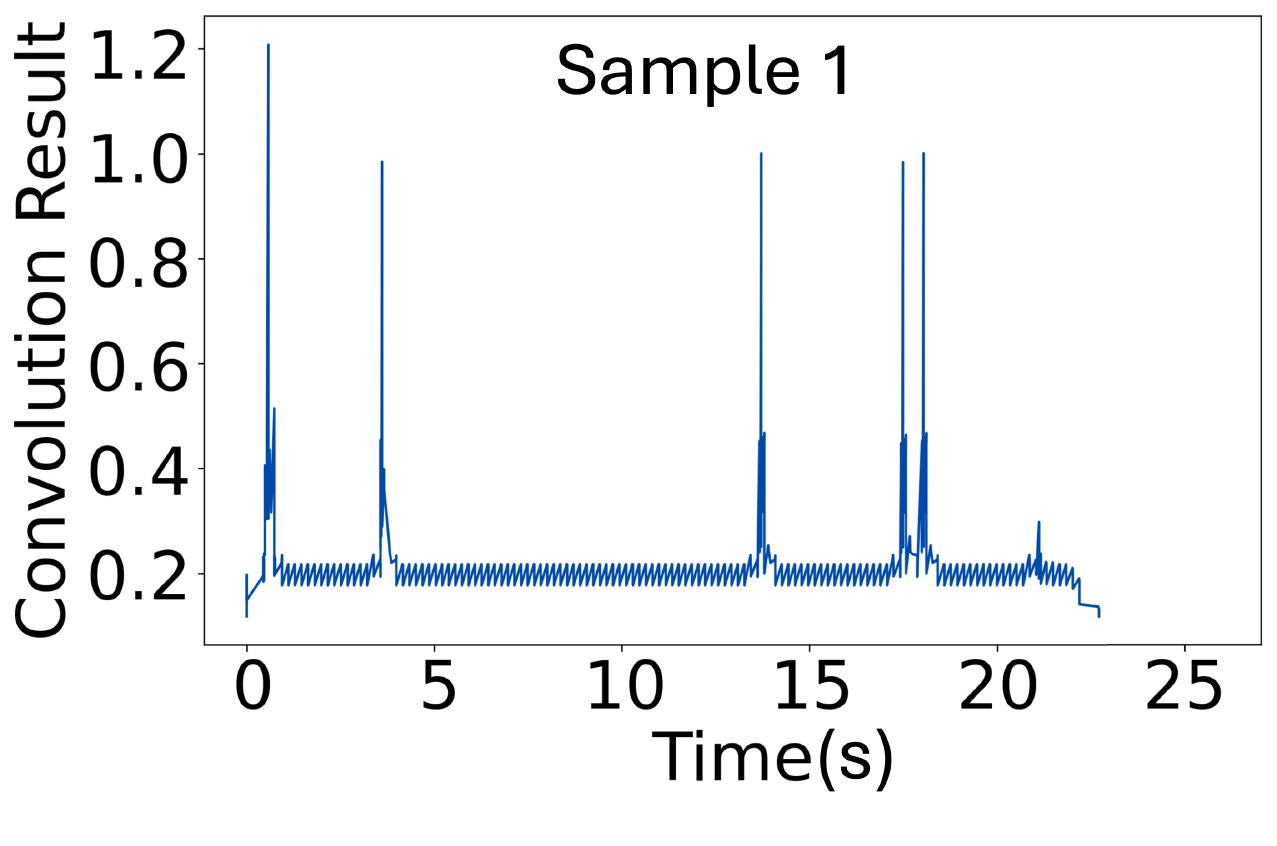}%
        \hfill
        \includegraphics[trim={3.3cm 1cm 0.1cm 0.1cm},clip,width=0.45\linewidth]{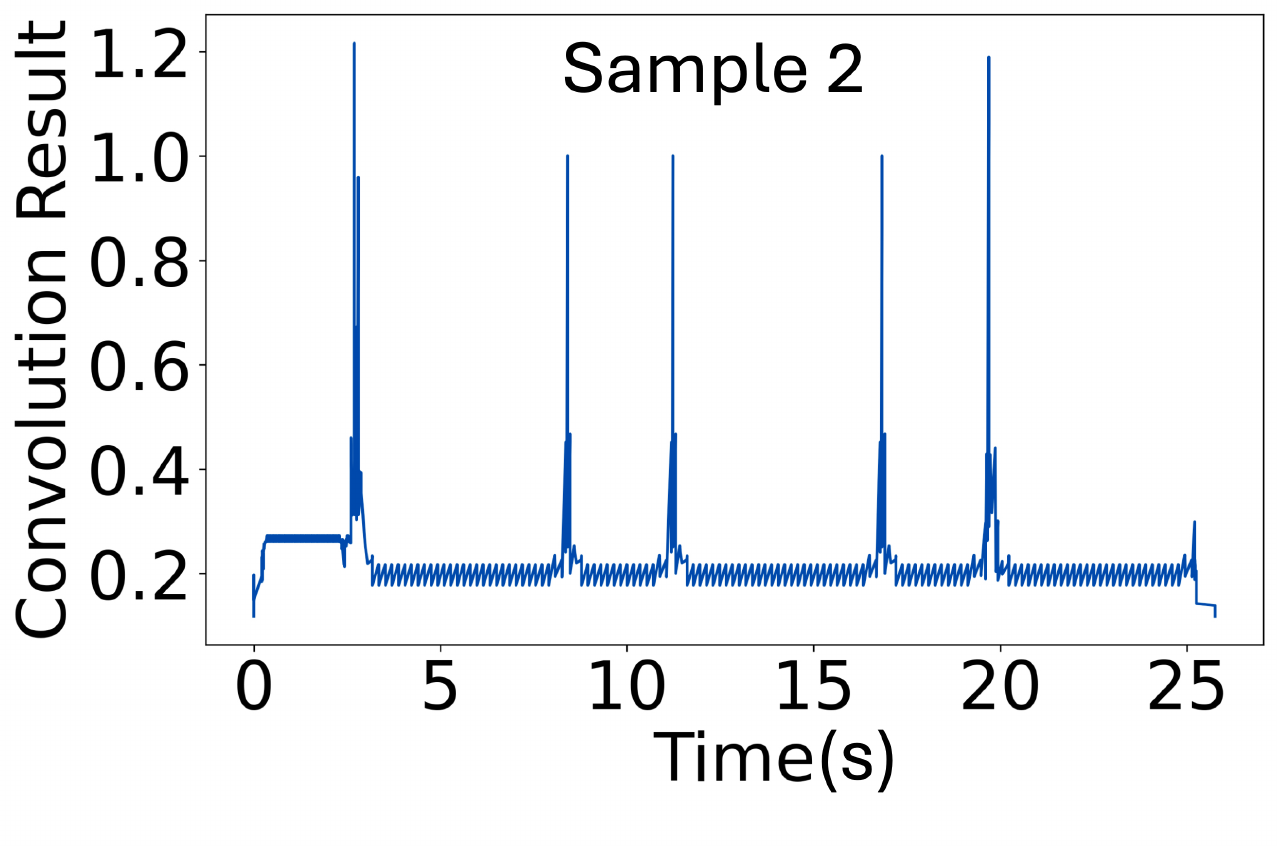}%
        \caption{Convolution for ``turn on switch''.}
        \label{fig:convolution_results}
    \end{subfigure}
    \hfill
    \begin{subfigure}[t]{0.5\linewidth}
        \centering
        \includegraphics[trim={0.1cm 1cm 0.1cm 0.1cm},clip, width=0.54\linewidth]{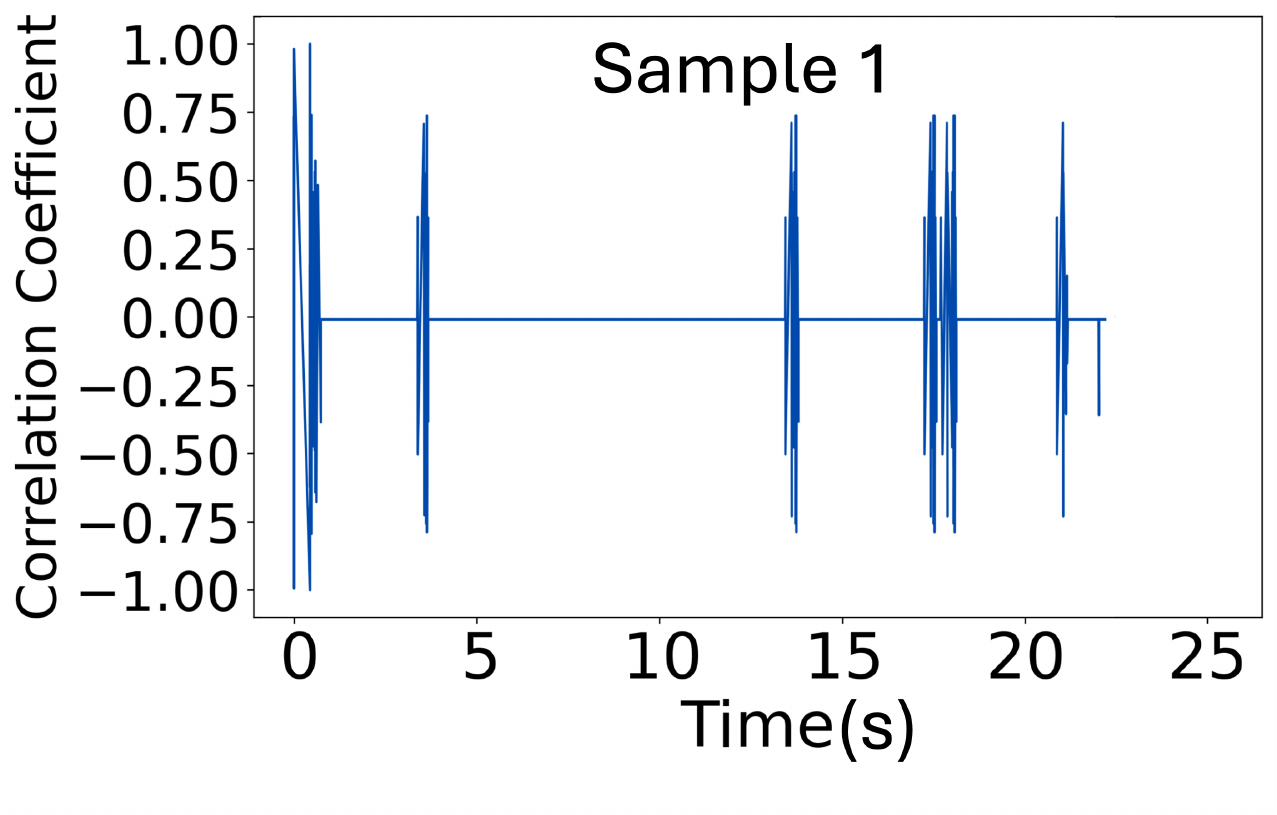}%
        \hfill
        \includegraphics[trim={4.2cm 1cm 0.1cm 0.1cm},clip,width=0.44\linewidth]{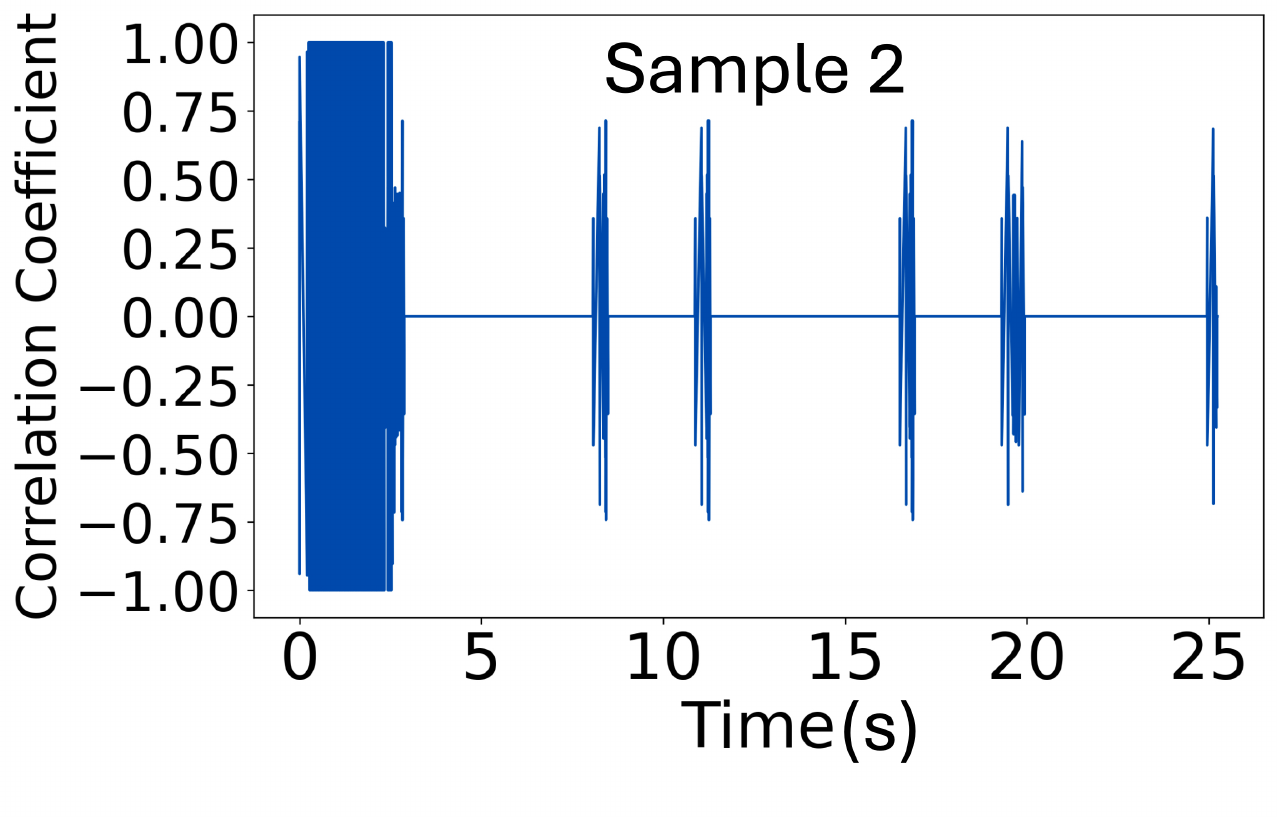}%
        \caption{Correlation for ``turn on switch''.}
        \label{fig:correlation_coefficient}
    \end{subfigure}
    \vspace{-0.2cm}
    \caption{Convolution and correlation results for two ``turn on switch'' samples.}
    \label{fig:combined_results}
    \vspace{-0.6cm}
\end{figure}

\mypara{Convolution-based statistics.} Consider the results of the convolution operation (Figure~\ref{fig:convolution_results}) applied to the two original ``turn on switch'' action traces depicted in Figure~\ref{fig:size-time-plot-comparison}. We can see that the convolution result is close to or above~1, meaning that the convolution operation is able to identify significant matches between the observed traffic pattern and the known command pattern. This indicates that the specific features or sequences in the traffic that correspond to the ``turn on switch'' action are effectively highlighted by the convolution process.

We note that we used a kernel designed to match the expected traffic pattern generated by the ``turn on switch'' action. The kernel was crafted based on the characteristic sequence of packet sizes and intervals that we observed in our preliminary analysis for this specific action. Later in Section~\ref{subsec:param_tuning}, we expand on the implications of selecting an adequate convolution kernel and of finding an adequate threshold for extracting command message information. By convolving this kernel with the observed traffic data, we sought to amplify parts of the signal that match the expected pattern. We determined a threshold to effectively identify instances where the convolution correctly identifies segments in the traffic data that correspond to the ``turn on switch'' action, based on the predefined kernel. The accurate identification of these segments is instrumental for reconstructing the robot's actions from the encrypted traffic. 

After obtaining the convolution results, we extract a set of statistics from the resulting signal to build a set of convolution-based features. Table~\ref{table:convolution-statistics} presents a summary of the features derived from our analysis. The example in Table~\ref{table:convolution-statistics} shows the features from the second ``turn on switch'' sample. We observe that there are six Cartesian command messages issued during the sample duration, as reflected in the total of six clusters detected. The table captures and shows the statistics relating to these commands. For example, on average, there is a gap of approximately 4 seconds between each cluster. Additionally, higher-level information such as the total time span and average time gap provides insights into the general dynamics of the action: how frequently the Cartesian command messages are issued and whether they are closely spaced. The maximum convolution value can offer an idea about the number of parameters specified in the Cartesian command message, as this typically results in larger feedback packets.

\begin{table*}[t]
\centering
\captionsetup{justification=centering} %
\resizebox{0.23\linewidth}{!}{
\begin{subtable}[t]{0.30\textwidth}
\centering
\scriptsize
\begin{tabularx}{\textwidth}{lr}
\toprule
\textbf{Metric} & \textbf{Value} \\
\midrule
Mean & 0.2454 \\
Standard Deviation & 0.096 \\
Median & 0.2172 \\
25th Percentile & 0.2163 \\
75th Percentile & 0.2616 \\
Maximum & 1.2160 \\
Minimum & 0.1181 \\
Skewness  & 6.0470 \\
Kurtosis & 48.9907 \\
Total Clusters & 6 \\
Total Time Span & 16.9991 \\
Average Time Gap & 3.3998 \\
\bottomrule
\end{tabularx}
\caption{Convolution features.}
\label{table:convolution-statistics}
\end{subtable}%
}
\hspace{1.5cm}
\resizebox{0.33\linewidth}{!}{
\begin{subtable}[t]{0.48\textwidth}
\centering
\scriptsize
\begin{tabularx}{\textwidth}{lr}
\toprule
\textbf{Metric} & \textbf{Value} \\
\midrule
Mean & -0.0019 \\
Standard Deviation & 0.5653 \\
Median & 0.0 \\
25th Percentile & 0.0 \\
75th Percentile & 5.3926e-17 \\
Maximum  & 1 \\
Minimum  & -1 \\
Skewness & 0.0039 \\
Kurtosis & -0.1747 \\
Number of Clusters & 1 \\
Total Length of Clusters & 2.0910 \\
Average Length of Clusters & 2.0910 \\
Total Time Span of Analysis & 2.0910 \\
Average Time Gap between Clusters & 0 \\
\bottomrule
\end{tabularx}
\caption{Correlation features.}
\label{table:correlation-coefficient-statistics}
\end{subtable}
}
\vspace{-0.2cm}
\caption{Convolution and correlation features for ``turn on switch'' (sample 2).}
\label{table:comparison-convolution-correlation}
\vspace{-0.6cm}
\end{table*}

\mypara{Correlation-based statistics.} Figure~\ref{fig:correlation_coefficient} depicts the results of the correlation coefficient operation for two original ``turn on switch'' action traces, previously shown in Figure~\ref{fig:size-time-plot-comparison}. We can observe a distinct difference in the traffic pattern of the two samples. The first sample does not exhibit any clusters, suggesting the absence of gripper speed commands, despite a high initial value. Conversely, the second sample displays a prominent cluster at the start, lasting  $\sim$2 seconds.

To enhance the accuracy of our detection, we establish a criterion where a cluster must consecutively last over 1 second to be considered a positive match. This threshold aligns with the inherent activity of gripper speed commands,  which typically persist for a certain duration to maintain a consistent speed or force applied by the gripper, thereby ensuring that we accurately identify when and for how long these commands are present in the action traffic.

Table~\ref{table:correlation-coefficient-statistics} provides a summary of the correlation coefficient statistics, offering details similar to those discussed in the convolution analysis. In this context, the cumulative and average lengths of clusters offer additional insights into the duration of commands. We employ the correlation coefficient to identify command messages that exhibit varying recurring patterns (e.g., the gripper speed command) as a means to reduce noise and false positives. By analyzing these statistics, we can, for instance, ascertain the duration of gripper movements.

\mypara{Our classifier's feature set.} The aforementioned convolution and correlation-based statistics comprise the main fuel for our traffic classifier. Specifically, the detection of each new command message allows us to refine the set of the above statistics with fine-grained information about commands' temporal dependencies. We obtain features for our model by processing these statistics into a summarized representation of messages' temporal dependencies. 

Concretely, our features include the \textit{average time between clusters}, shedding light on the intervals between command groups; \textit{total number of clusters}, quantifying the diversity of patterns; \textit{total length of all clusters}, giving an aggregate duration of detected patterns; and \textit{average length of clusters}, indicating the typical duration of individual patterns.  The \textit{time gaps between consecutive clusters} highlight intervals between patterns, while \textit{skewness and kurtosis} offer insights into their distribution. The \textit{total time span of clusters} provides data about the start of the first to the end of the last cluster, and the \textit{average time gap between clusters} showcases the average intervals between these occurrences. 
Collectively, these features paint a detailed picture of the robot's operational patterns, capturing not only the frequency and duration of command messages but also transitions and relationships within them. 
In addition, we also include in our final feature set a collection of common summary statistics extracted from network traffic (akin to k-FP~\cite{kfingerprinting}) and that relate to the timing and volume of communication (e.g., number of packets exchanged, or average inter-packet timing).

\subsection{Convolution Parameters}
\label{subsec:param_tuning}

The results of our convolution operation are mostly guided by the choice of two parameters: the convolution kernel, and a threshold for determining the occurrence of a command message. Below, we detail our choice for both parameters.

\mypara{Convolution kernel.} In our experiments, we find that the specific choice of the convolution kernel—a pattern or template used to detect similar patterns in the observed data—does not significantly impact the accuracy of the results. Essentially, a kernel represents the overall shape or signature of a particular type of traffic pattern, such as a Cartesian or gripper position command message. For example, a kernel might be derived from a typical pattern of packet sizes and intervals observed for a specific robot command. We test this by using 10 different kernels for each type of command message, each extracted from 10 distinct samples of that command. Despite these variations, the accuracy of our traffic pattern detection remains consistent. This suggests that our method is robust to variations in the kernel, capable of accurately identifying traffic patterns regardless of slight differences in the kernel's shape.

\mypara{Convolution threshold.} Figure \ref{fig:accuracy-threshold} shows the variation in our classifier's accuracy across different convolution threshold (\textit{t}) values. The threshold is based on the range of convolution results obtained when the normalized signal is convolved with the normalized kernel, using the normalization factor derived from the kernel itself. A threshold value of 0 means that every positive convolution result, no matter how small, is classified as a Cartesian command message. This can lead to a high rate of false positives, as even minimal similarities between the observed traffic and the kernel are flagged as matches. Conversely, a threshold of 1.3 sets a very high bar for detection, meaning that only very strong matches—those with convolution results exceeding this value—are considered true positives. This strict criterion can increase the likelihood of false negatives, as it may overlook less pronounced but still relevant patterns. In our analysis, we find that a threshold value of approximately 0.9 strikes an optimal balance. 

\subsection{Evaluation of our Approach}
\label{subsec:sp_eval}

\begin{figure}[t]
    \centering
    \subfloat[Acc. for threshold \textit{t}.]{%
        \includegraphics[trim={0cm 0cm 0cm 0cm},clip,width=0.28\linewidth]{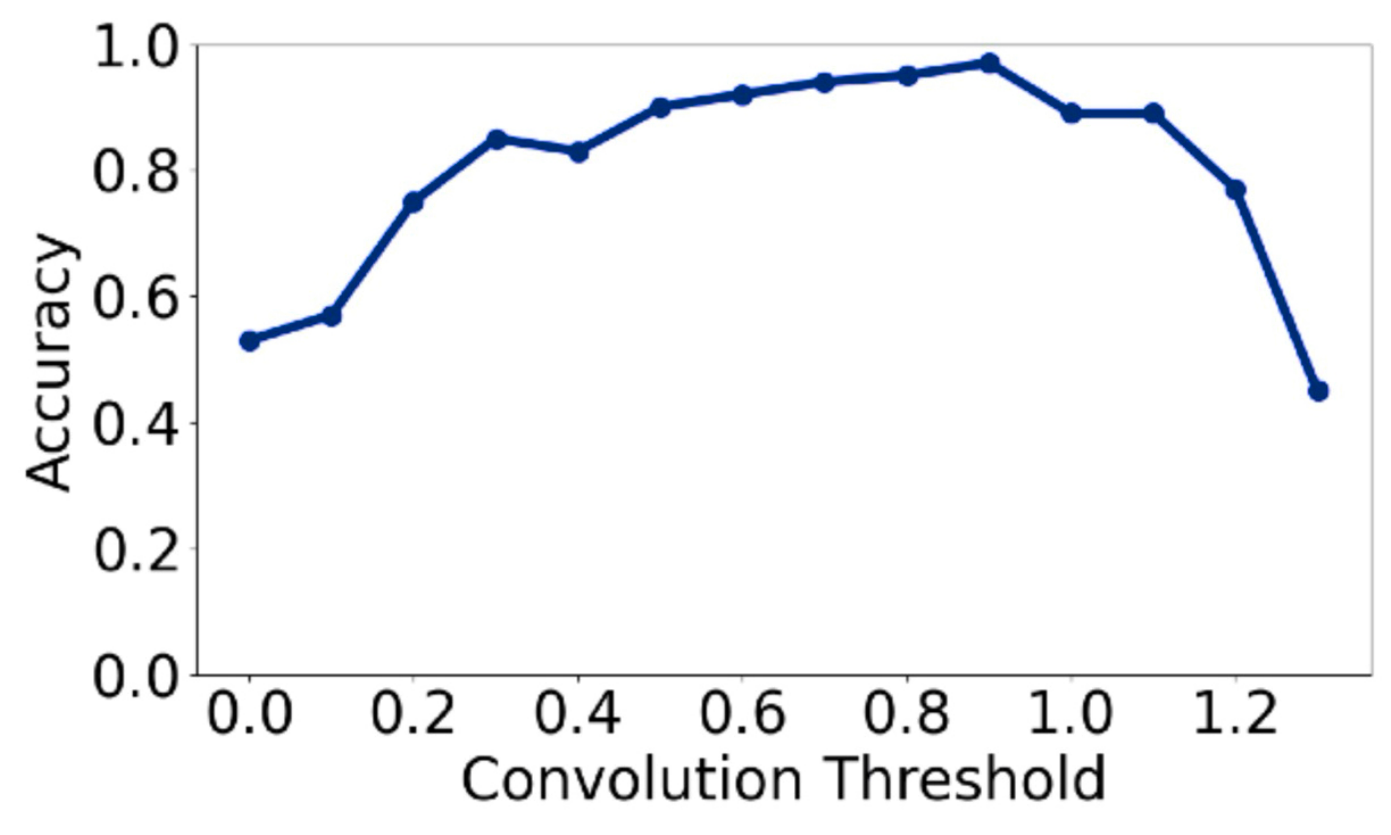}%
        \label{fig:accuracy-threshold}
    } 
    \subfloat[Conf. mat. ($t=0.9$).]{%
        \includegraphics[trim={0.1cm 0.1cm 0.1cm 0.65cm},clip,width=0.28\linewidth]{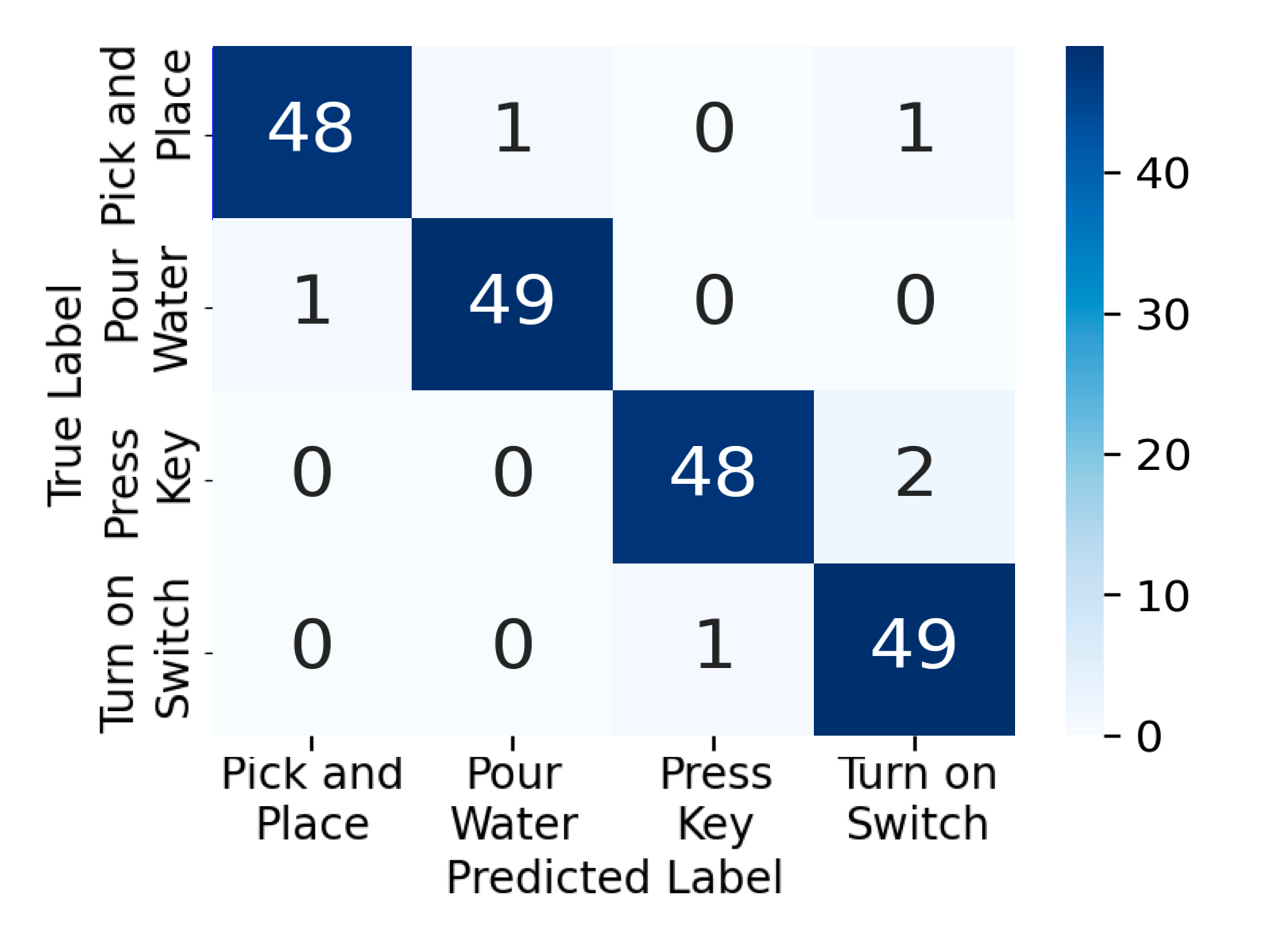}%
        \label{fig:confusion-matrix}
    }
    \subfloat[Top 20 features ($t=0.9$).]{%
        \includegraphics[trim={2cm 0cm 0cm 0cm},clip,width=0.4\linewidth]{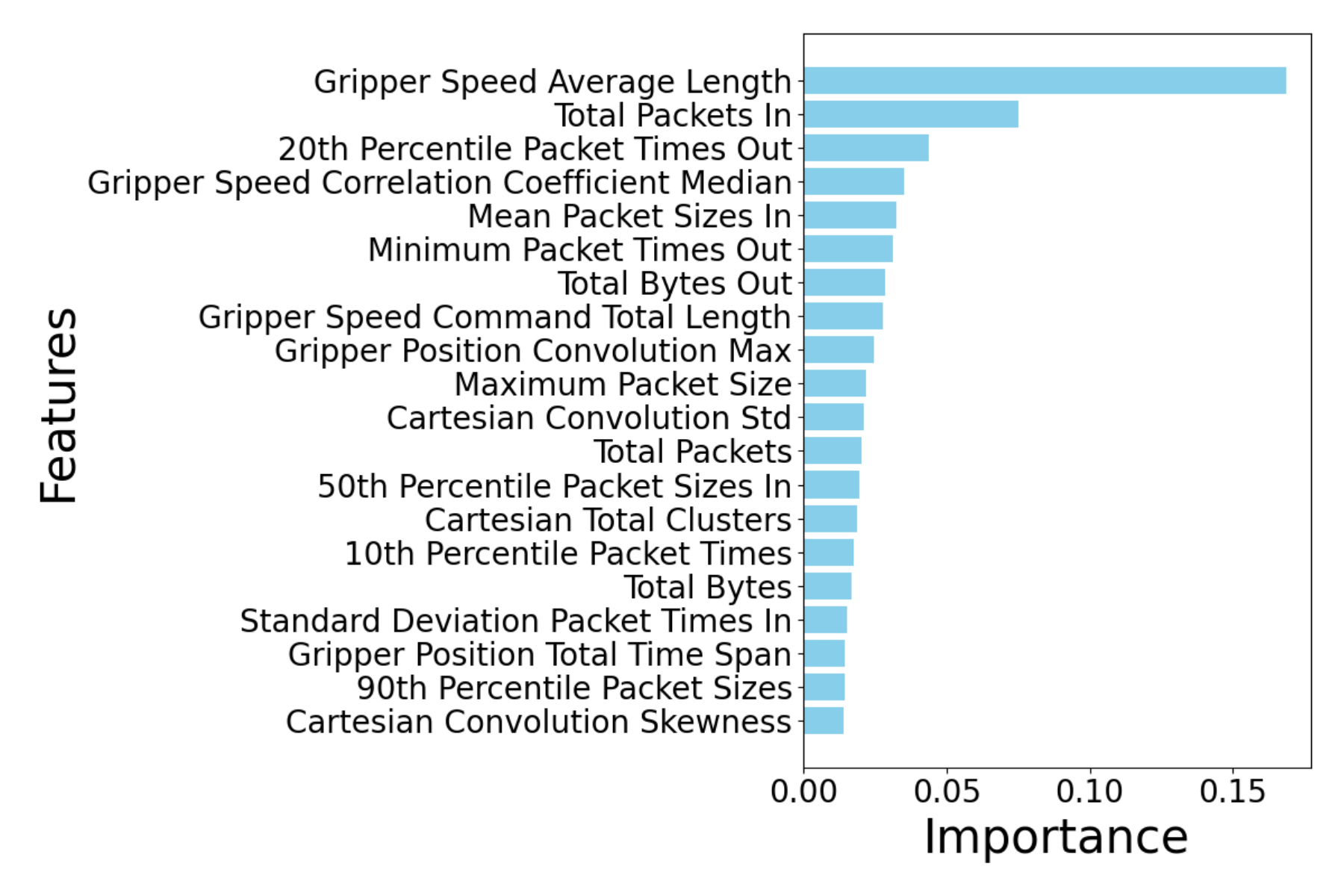}%
        \label{fig:feature-importance}
    }
    \vspace{-0.2cm}
    \caption{Classification results.}
    \vspace{-0.5cm}
\end{figure}

Our experiments reveal that our traffic analysis pipeline achieves an accuracy of 97\% in the robot actions dataset, thus showcasing the potential for network adversaries to compromise the privacy of users through the accurate recognition of robot activities.
Figure~\ref{fig:confusion-matrix} depicts the confusion matrix of our classifier on the robot's actions dataset, allowing for two interesting observations. First, we can see that the classifier occasionally mispredicts the ``pick and place'' and ``pour water'' with one another. This misclassification likely stems from the similar sequence of movements shared by these actions, which  involve the gripper's opening and closing motions used for grabbing and placing an object.  Second, actions that involve tapping motions, such as pressing a key and toggling a switch, are also occasionally misclassified as each other. Indeed, both classes exhibit quick and sharp motion patterns, which are typical of tapping actions, and thus distinct from the more fluid and prolonged actions of picking, placing, and pouring. Our classifier is able to leverage these unique patterns to more accurately differentiate between these two sets of actions.

\mypara{Feature importance.}
The top 20 most important features for our classifier's performance (depicted in Figure~\ref{fig:feature-importance}) comprise a mix of signal processing and summary statistics elements. 
Feature importance is determined based on the average \textit{gain}, which measures the improvement in the loss function when a feature is used for splitting in the XGBoost model.   From the signal processing category, the gripper speed average length emerges as the most significant feature, followed by the gripper speed command correlation coefficient median, ranking as the fourth most important, highlighting the role of gripper movement clusters in the classification. On the other hand, from the summary statistics category, the total number of incoming packets and the 20th percentile of outgoing packet inter-arrival times stand out as the second and third most important features, respectively. This highlights that different classes exhibit variations not only in the total number of packets sent but also in the timing between these packets. Such distinctions enable our classifier to correctly distinguish between classes. 

\section{Towards Efficient Defenses}
\label{sec:defenses}

Here, we explore different defense mechanisms and determine their effectiveness in protecting robotic operations from traffic analysis attacks. We evaluate two defenses: one based on the simple \textit{padding of packet sizes}, and a more complex one developed by ourselves, which we named \textit{latency-aware traffic modulation}.

At the core of the defense mechanisms we consider lies a concern over how these defenses may impact the correct and timely operation of the robot. As it stands, the robot can only proceed with another action after it has received an entire command message and feedback messages about that same command (both possibly segmented over multiple packets). Delaying this cycle would lead to delays in action execution. 
To ensure that action correctness and operational efficiency are maintained, the content within the command and feedback messages must be sequentially delivered before any robotic action can be executed. Moreover, the messages exchanged between the robot and the controller while a command's execution is ongoing should also not be delayed.

\subsection{Considered Defenses} 

\mypara{Padding packet sizes.} This defense obfuscates the size of packets, making it more challenging for an adversary to discern patterns based on packet size. This method does not introduce additional latency, as it does not affect the transmission time of the packets. However, it does result in additional bandwidth overhead due to the transmission of larger-sized packets.
The defense mechanism rounds up each packet size to the nearest multiple of $x$ * 100 bytes. We assume $x$ to be an integer ranging from 1 to 10 since the maximum packet size observed in our traces was under 1000 bytes. We also assume padding to occur only up to the maximum MTU size of 1500 bytes. For example, a packet of size 360 bytes would be padded to 400 bytes if $x=2$ (nearest multiple of 200) and to 500 bytes if $x=5$, while a packet of size 960 bytes would be padded to 1500 bytes if $x=8$.

\mypara{Latency-aware traffic modulation.} This defense strategy is inspired by constant-rate traffic analysis defenses (in the likes of Tamaraw~\cite{tamaraw}), which involve the use of a fixed packet size for all messages. Generically, applying such a defense to our setting would cause the segmentation of any robot command or feedback message larger than this fixed size into several packets of equal size, or, alternatively, the padding of the contents of small messages to meet this fixed packet size. Then, all packets would be sent at a specific and fixed time interval.

However, the above method can introduce delays, particularly if the frequency of packet transmission does not align with the speed at which the robot's controller operates. Concretely, the Kinova Gen3 robotic arm features a closed-loop control system operating at a frequency of 1 kHz. Our analysis of the robot operation traffic revealed that the command message type with the \textit{shortest inter-packet intervals} is the speed command, which averages about 80 packets per second. Any latency in a single command/feedback message that results in the message arriving noticeably late to the controller is unacceptable. 

To address this, we developed a \textit{latency-aware traffic modulation} scheme that helps maintain latency within an acceptable range while adequately obfuscating traffic patterns. Like constant-rate defenses, our technique still relies on sending dummy packets to fulfill fixed packet sending rates and padding short packets to meet the pre-configured fixed packet size. However, we leave packets longer than this fixed size (and which cannot be broken down without exceeding the acceptable latency range) at their original size. In turn, packets that can be segmented into multiple smaller packets without resulting in unacceptable delays are appropriately split. This approach still introduces some latency, but can keep it within an acceptable range for the robotic arm's operation~\cite{kinova3}.

As a concrete formulation, let $s_o$ be the original packet size, and let $s_p$ be the predefined padded packet size. Let $L$ be the permissible latency 0.001 second in the case of running the Kinova controller and $t_i$ be the chosen time interval for transmitting each packet. The calculated packet size $s_c$ and the number of segments $n$ into which the original packet is divided can be defined as follows:

\begin{minipage}{0.45\textwidth}
{\footnotesize
\begin{equation}
    s_c = 
    \begin{cases} 
      s_p & \text{if } s_o \leq s_p, \\
      \frac{s_o}{\left\lfloor \frac{L}{t_i} \right\rfloor} & \text{if } \left\lceil \frac{s_o}{s_p} \right\rceil \cdot t_i > L, \\
      s_p & \text{otherwise}.
    \end{cases}
\end{equation}
}
\end{minipage}
\hfill
\begin{minipage}{0.45\textwidth}
{\footnotesize
\begin{equation}
    n = 
    \begin{cases} 
      1 & \text{if } s_o \leq s_p, \\
      \left\lfloor \frac{L}{t_i} \right\rfloor & \text{if } \left\lceil \frac{s_o}{s_p} \right\rceil \cdot t_i > L, \\
      \left\lceil \frac{s_o}{s_p} \right\rceil & \text{otherwise}.
    \end{cases}
\end{equation}
}
\end{minipage}

\begin{figure}[b]
\vspace{-0.4cm}
	\centering
	\subfloat[Classifier's accuracy.]{\includegraphics[trim={0cm 0cm 0cm 0cm},clip,width=0.35\linewidth]{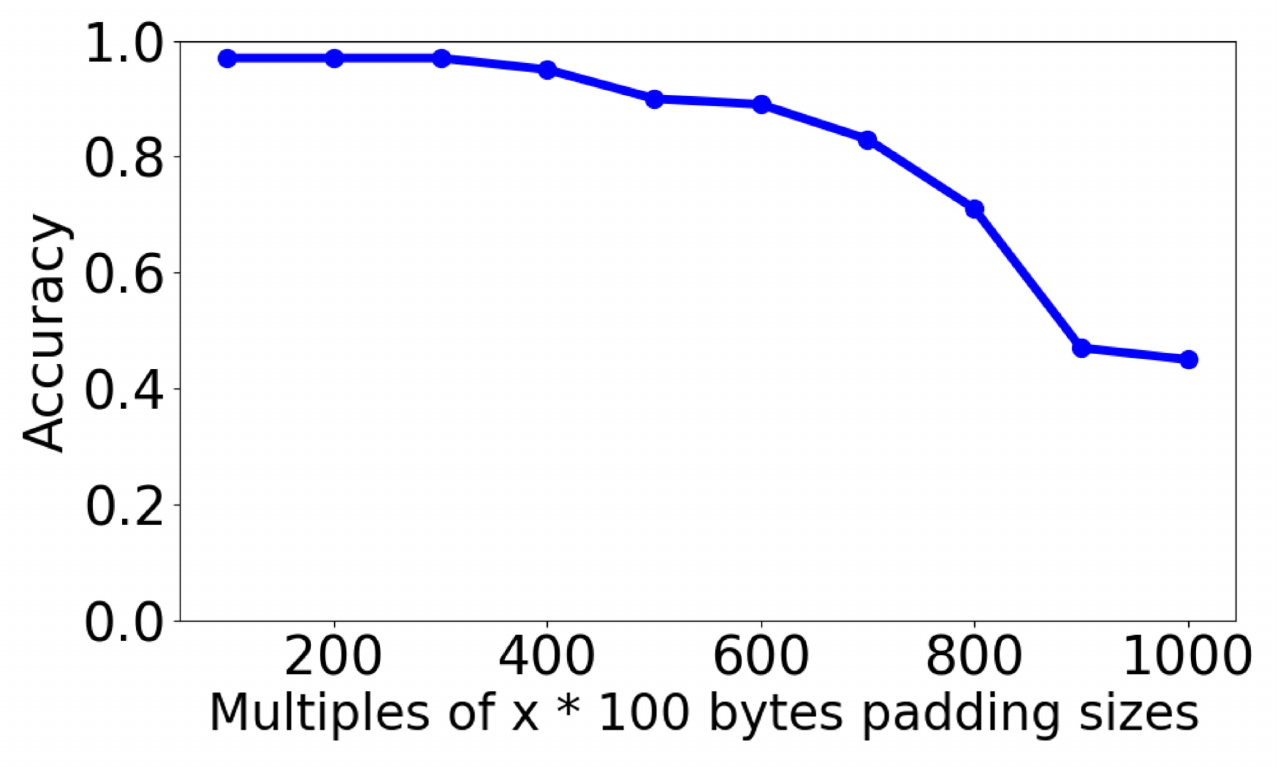}\label{fig:accuracy-size}} \hspace{0.15\linewidth}
	\subfloat[Bandwidth overhead.]{\includegraphics[trim={0cm 0cm 0cm 0cm},clip,width=0.35\linewidth]{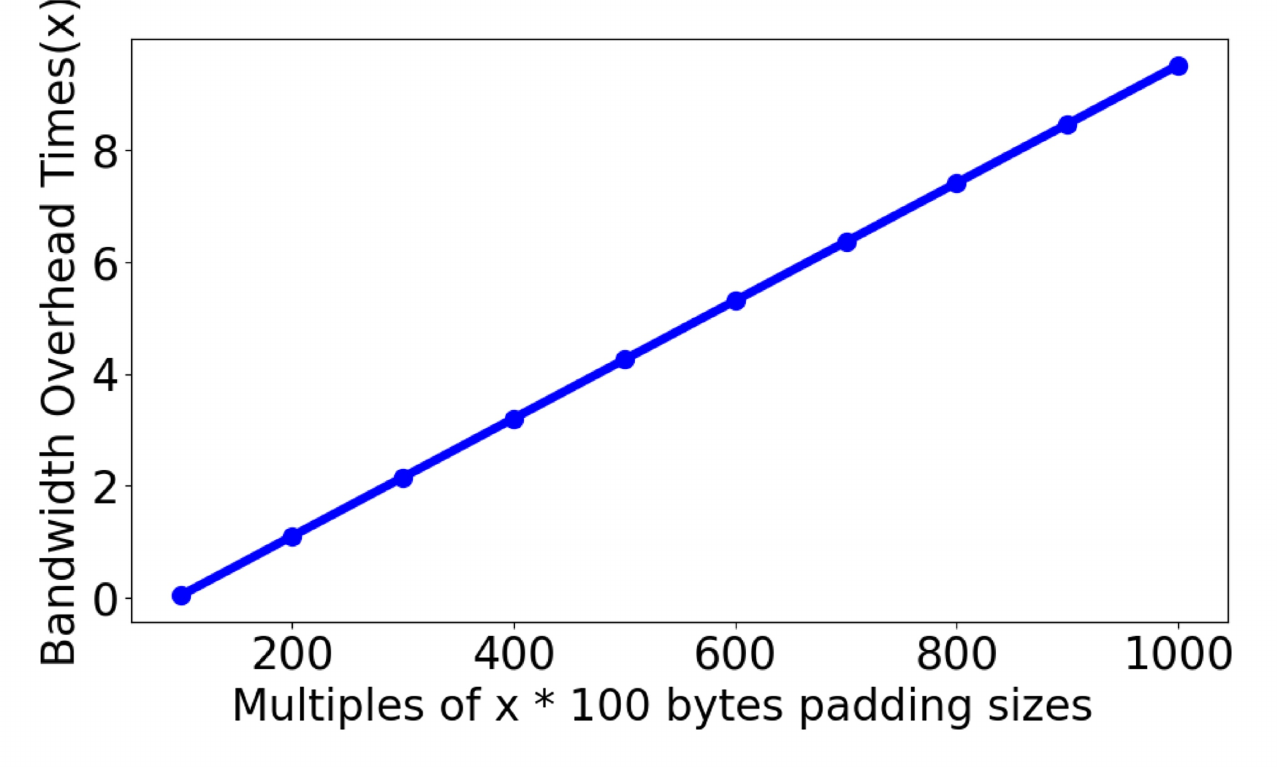}\label{fig:overhead-size}}
 \vspace{-0.2cm}
	\caption{Results for the padding-only defense.}
	\label{fig:padding}
\end{figure}

\subsection{Evaluation of Defenses}

\mypara{Padding-only defense.} Figure~\ref{fig:accuracy-size} illustrates the relationship between the accuracy of our classifier and the size of padded packets when the simpler padding-only defense is employed. The classifier retains a relatively high accuracy (above 80\%) for packets padded to the nearest multiple of 700 bytes ($x=7$), decreasing its accuracy to $\sim$40\% when packets are padded to the nearest multiple of 1000 bytes ($x=10$). For analyzing the security/efficiency trade-off in more detail, Figure~\ref{fig:overhead-size} depicts the average bandwidth overhead percentage of the defense when considering our robot actions dataset. As observed, the overhead percentage increases linearly with the padding size, reaching approximately 700\% overhead when the padding size is set to the nearest multiple of 800 bytes ($x=8$). At this padding size, the accuracy of our classifier is still relatively high ($\sim$70\%). Thus, we conclude that this defense is largely ineffective for low padding sizes, achieving only a moderate protection for a large bandwidth overhead. 

\mypara{Latency-aware traffic modulation defense.} For evaluating this defense, we perform a set of experiments with different permissible latency: 0.01s, matching the fastest average command speed, specifically for speed commands; 0.001s, aligning with the controller's operational frequency, and; 0.0001s, the frequency allowing the transmission of the shortest command/feedback message. While the speed command typically exhibits the fastest average frequency, some command messages contain packets with significantly smaller single inter-packet intervals.

\begin{figure}[t]
	\centering
	\subfloat[Classifier's accuracy.]{\includegraphics[trim={0cm 0cm 0cm 0cm},clip,width=0.48\linewidth]{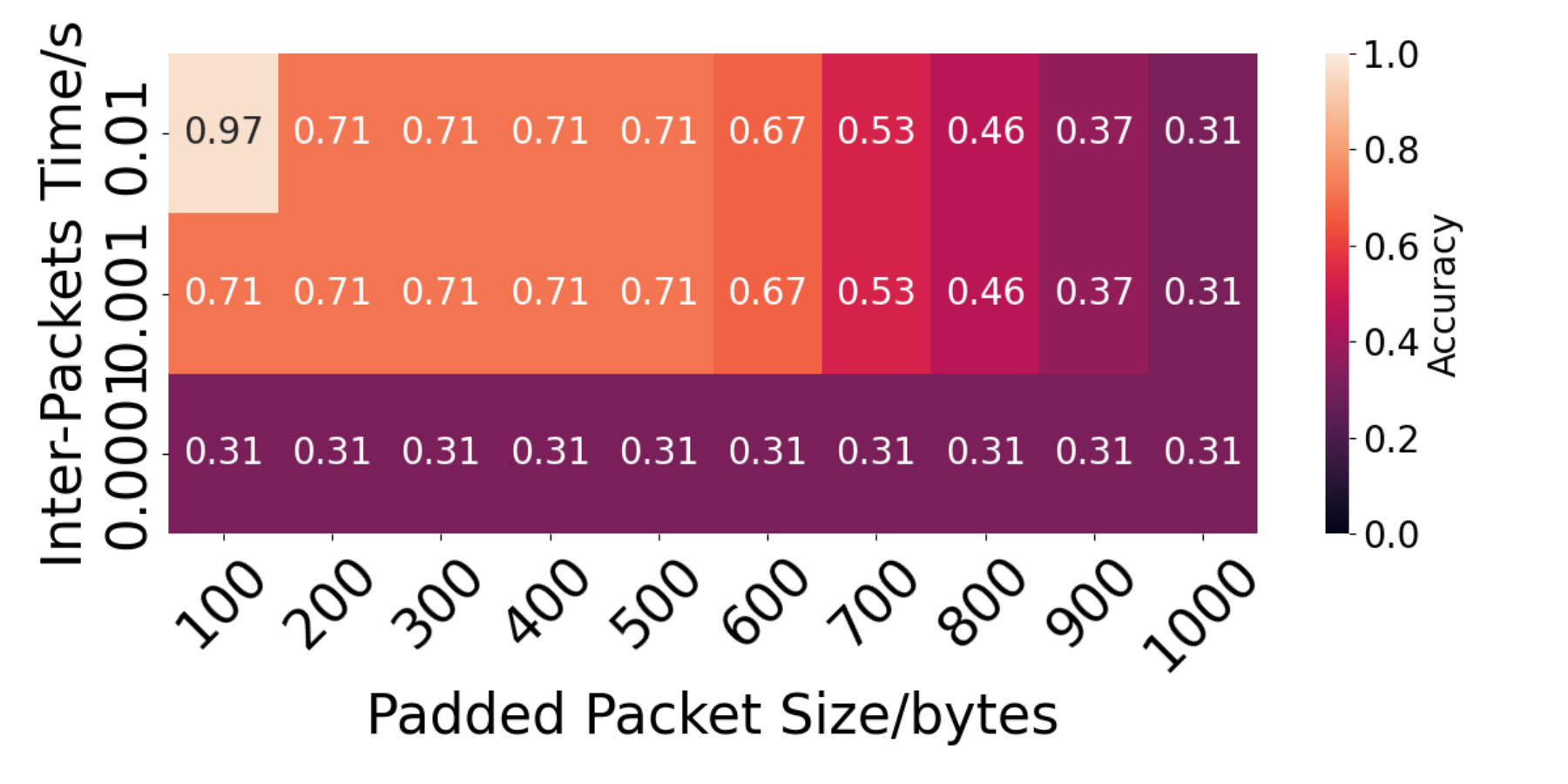}\label{fig:accuracy-size-time}} 
	\subfloat[Bandwidth overhead.]{\includegraphics[trim={0cm 0cm 0cm 0cm},clip,width=0.48\linewidth]{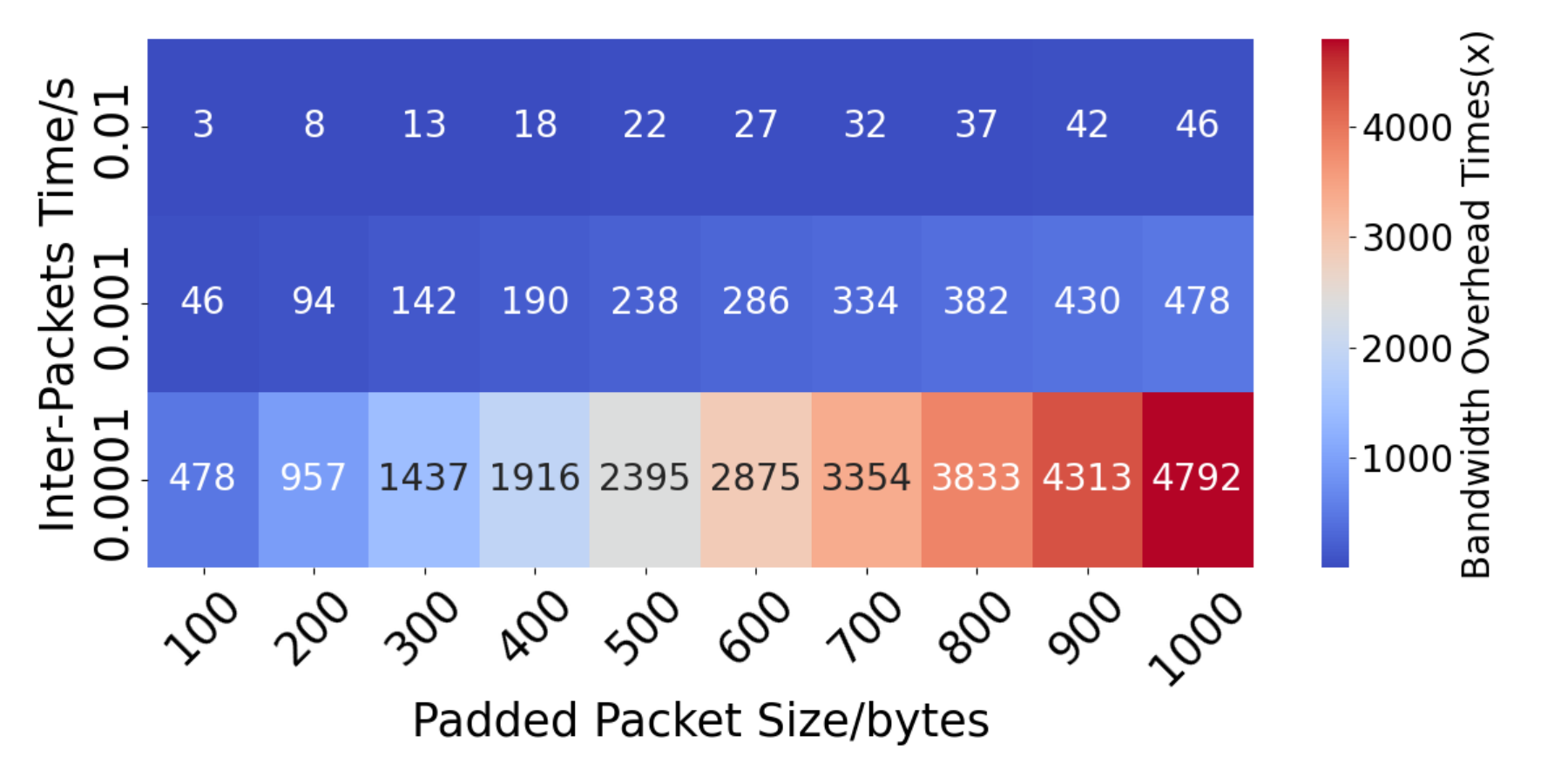}\label{fig:overhead-size-time}}
 \vspace{-0.3cm}
	\caption{Results for the latency-aware defense.}
	\label{fig:latency-aware}
 \vspace{-0.4cm}
\end{figure}

Figure~\ref{fig:accuracy-size-time} illustrates the relationship between the effectiveness of our defense mechanism and the degree of packet padding and segmentation applied to robot traffic. We can see that, for fixed packet sending rates of 0.01s and 0.001s, the defense is only reasonably effective once packets are also padded to a large size (900 bytes for degrading classification accuracy below 40\%). However, we can also see that, for all the padding sizes under test, sending packets at a fixed frequency of 0.0001s is sufficient for reducing the accuracy of the classifier to approximately 30\%. 
Figure~\ref{fig:accuracy-size-time} also suggests that an adversary may still be able to infer information about the actions being performed (accuracy of $\sim$30\%) due to the total time duration of each action (which we do not obfuscate).

Figure~\ref{fig:overhead-size-time} shows the bandwidth overhead incurred by the latency-aware traffic modulation defense. Bandwidth overheads increase both with the packet sizes and with the packet sending frequency used in the defense. The figure suggests that the latter cause more pronounced bandwidth overhead, given the additional packets exchanged per time unit. When sending packets at a 0.01s rate, the defense reduces our attack's accuracy to 31\% but incurs a staggering 478 times bandwidth overhead. Unfortunately, this trade-off is not practical in realistic settings, indicating a pressing need for the development of more sophisticated low-delay defense strategies that can provide robust privacy protections. Reasoning about such defenses is a compelling direction for future work.

\section{Limitations and Future Work}

\mypara{Extension to the open-world.} Our evaluation matches that of a closed-world attack~\cite{shen2023subverting} where we attempt to distinguish between a restricted set of four different actions performed by the robot. In the future, we aim to analyze the effectiveness of our approach in the open-world setting~\cite{OWWF}, where the robot is allowed to perform a range of tasks beyond those the adversary seeks to identify.

\mypara{Traffic analysis under varying network conditions.} We assumed a stable and high-performance network connection that was free from interference (e.g., packet drops, jitter, etc., caused by network malfunctions). An interesting direction for future work would be to consider the traffic analysis of robot traffic under varying network conditions. Different network environments may introduce unique challenges and opportunities for attack and defense strategies~\cite{realisticWF}.

\mypara{Measuring robot traffic information leakage.} We aim to explore more systematic methods for evaluating information leakage in robot traffic~\cite{wfes,wefde,deepse}. Automating a security evaluation pipeline for robotics applications could  enhance the overall effectiveness of security assessments applied to these systems.

\section{Related Work}
\label{sec:rw}

The fingerprinting scenarios we enumerate below target the re-identification of traffic that follows relatively stable patterns. We argue that fingerprinting robot actions presents unique challenges -- complex robot activities might involve a variable number of smaller operations which can be performed in different orders and  last for different amounts of time. 
In our work, we rely on a new set of manually-engineered features based on signal processing operations, combining these with traffic statistics that have been used before for fingerprinting purposes.

\mypara{Website fingerprinting.}
Despite the use of encrypted tunnels like VPNs and low-latency anonymity networks such as Tor, web traffic retains characteristic traffic patterns that allow eavesdroppers to map those patterns to specific websites~\cite{kfingerprinting}. While the creation of these mappings has traditionally resorted to manual feature engineering and classical machine learning algorithms~\cite{kfingerprinting,CUMUL}, recent developments rely on the use of deep learning to improve the accuracy of website fingerprinting attacks~\cite{rahman2019tik,sirinam2018deep,shen2023subverting}. There also exists prolific literature on defenses~\cite{mathews2022sok}, ranging from rather secure but inefficient constant-rate padding defenses~\cite{tamaraw}, to more efficient defenses like FRONT~\cite{gong2020zero} or RegulaTor~\cite{regulator}.

\mypara{Video fingerprinting.}
Video streaming protocols that adjust their bit rates in response to network conditions manifest traffic patterns that fluctuate based on the video content and its resolution, enabling the development of fingerprinting techniques which can identify a given video being streamed, even if it is encrypted~\cite{beautyBurst,netflix}. 
Proposed defenses include the generation of synthetic video flows that shield the videos being watched by users~\cite{smaug}, or noise addition~\cite{smartswitch}.

\mypara{IoT device fingerprinting.}
IoT devices typically exhibit regular and predictable traffic patterns which are only partially shrouded by encryption, enabling eavesdroppers to infer which specific devices operate within a smart-household~\cite{smartHome}. 
Similarly to website fingerprinting, different approaches for IoT device fingerprinting have focused both on the collection of summary statistics from IoT device traffic~\cite{chowdhury2020network,engel2019}, or on the analysis of network traces with deep neural networks~\cite{dong2020your}. Existing defenses include improved padding schemes~\cite{engelberg2022classification} and the perturbation of traffic flows through deep learning methods~\cite{shenoi2023ipet}.

\section{Conclusion}

In this paper, we investigated encrypted traffic analysis in robotics, emphasizing its implications for sensitive applications (e.g., domestic, healthcare). Leveraging machine learning classifiers fueled with signal-based processing features, we were able to discern robotic actions from encrypted traffic patterns, shedding light on significant privacy concerns. Our findings highlight potential vulnerabilities within robotic communication channels, calling for the development of enhanced security measures as the deployment of collaborative robots becomes widespread.  Moreover, while we focused on collaborative robots, our method and findings likely also apply to industrial robot arms, such as those used in manufacturing. Their repetitive, pre-configured actions may make them even more vulnerable to traffic analysis attacks, casting the need for more advanced defense mechanisms.

\bibliographystyle{splncs04}
\bibliography{main}

\end{document}